\documentclass[aps,prl,twocolumn,showpacs,superscriptaddress,groupedaddress]{revtex4-2}  
\usepackage[T1]{fontenc}
\usepackage{graphicx}  
\usepackage{dcolumn}   
\usepackage{bm}        
\usepackage{amssymb}   
\usepackage{amsmath}
\usepackage{bbold}
\usepackage{array}
\usepackage{makecell}
\usepackage{braket}
\usepackage{titlesec} 
\usepackage[colorlinks,citecolor=red,urlcolor=blue,hypertexnames=true]{hyperref}
\usepackage{subfigure}

\usepackage[usenames,dvipsnames,svgnames,table]{xcolor} 

\begin{document}

\widetext


\title{\textcolor{Sepia}{\textbf{\Large Wormhole calculus without averaging \\  from \\ $O(N)^{q-1} $ tensor model}}}
\author{{Sayantan Choudhury${}^{1,2*}$},\thanks{{\it Corresponding author,}\\
	{{ E-mail:sayantan.choudhury@niser.ac.in,  sayanphysicsisi@gmail.com}}}~ K. Shirish ${}^{3}$}
~~~~~~~~	\\


\affiliation{${}^{1}$National Institute of Science Education and Research, Jatni, Bhubaneswar, Odisha - 752050, India.}
\affiliation{${}^{2}$Homi Bhabha National Institute, Training School Complex, Anushakti Nagar, Mumbai - 400085,
	India.}
\affiliation{${}^{3}$Visvesvaraya National Institute of Technology, Nagpur, Maharashtra, 440010, India}

\begin{abstract}
The SYK model has a wormhole-like solution after averaging over the fermionic couplings in the nearly $AdS_2$ space.  Even when the couplings are fixed the contribution of these wormholes continues to exist and new saddle points appear which are interpreted as "half-wormholes". In this paper, we will study the fate of these wormholes in a model without quenched disorder namely a tensor model with $O(N)^{q-1}$ gauge symmetry whose correlation function and thermodynamics in the large $N$ limit are the same as that of the SYK model. We will restate the factorization problem linked with the wormhole threaded Wilson operator, in terms of global charges or non-trivial cobordism classes associated with disconnected wormholes. Therefore for the partition function to factorize especially at short distances, there must exist certain topological defects which break the global symmetry associated with wormholes and make the theory devoid of global symmetries. We will interpret these wormholes with added topological defects as our "half-wormholes". We will also comment on the late time behavior of the spectral form factor,  particularly its leading and sub-leading order contributions coming from higher genus wormholes in the gravitational sector. Finally we will show how, the other non-trivial saddles from "half-wormhole" dominate and give rise to unusual thermodynamics in the bulk sector due to non-perturbative effects.

\end{abstract}

\pacs{}
\maketitle
\section{\textcolor{Sepia}{\textbf{ \Large Introduction}}}
\label{sec:introduction}
In the context of AdS/CFT correspondence wormholes play significant role in understanding the physics of quantum black holes, the long-time behavior of spectral form factor and Hawking-Bekenstein entropy \cite{Page:1993wv,Almheiri:2019qdq,Almheiri:2019hni}. Recent developments has uncovered many interesting links between wormholes and random matrix theory and even humanly traversable wormhole has been studied \cite{Maldacena:2018lmt,Penington:2019kki,Maldacena:2017axo,Maldacena:2004rf,Marolf:2021kjc,Johnson:2021rsh,Stanford:2020wkf}.

But despite of this, the addition of wormhole leads to puzzles such as the factorization problem i.e the partition function of the combined system L and R from the boundary perspective factorizes. In other words, they are just the product of two boundary components as $Z_{LR} = Z_{L}Z_{R}$, but in the bulk computation, the contribution of the wormhole tends to spoil the factorization which gives rise to the factorization puzzle.

However, the factorization puzzle isn't a paradox since these wormhole contributions are sub-dominant or suffer from instabilities in all UV complete string theories. But while studying the spectral form factor the long time behavior of the linear ramp is described by wormholes and if do not perform the time averaging in the context of the SYK, the long time behavior of the spectral form factor has large oscillations, close to the size of the ramp. Therefore one might suspect that there are wormhole contributions that describe these oscillations and must be taken into account.

Recently in a paper by {\it Saad, Shenker, Stanford, and Yao} \cite{Saad:2021rcu, kk,ramsey,philsaadtalk,LR} where they studied the SYK model with fixed coupling and computed the observables $z_{L}z_{R}$ and the correlation between the two boundary systems $L \text{and} R$ by introducing the collective field variables $G_{LR}, \Sigma_{LR}$. The model has wormhole solutions when $G_{LR}, \Sigma_{LR} \not = 0$. while studying this model they found two different types of saddle points one describing the wormhole and another one which they called "half-wormhole". This new half-wormhole contribution depends strongly on the coupling and comes from the no self-averaging part of the theory and vanishes when the average is considered $\langle z_{L}z_{R} \rangle $. However, when the contribution of both the wormhole and half-wormhole are taken into account the factorization of partition function seems to be restored in the model with fixed coupling. 

In this paper, we study an SYK related quantum mechanical model without the quenched disorder in which the complex gauged fermionic  fields transform in the fundamental representation of $q-1$ copies of $O(N)$ \cite{Sachdev:1992fk,Hosur:2015ylk,Cao:2021upq}. It has been shown that model behaves similar to that of SYK in the large $N$ limit even though they differ at finite values of $N$. In the large $N$ limit by integrating out the fermionic fields we found that the sub-dominant saddle points of gauge holonomies play a crucial role in unitarizing the black hole physics, therefore these saddle points are similar to  half-wormhole saddles in the SYK.

We then construct a wormhole threaded Wilson operator in the bulk connecting the two boundaries and will restate the problem of factorization of bulk field into left and right CFT operator into the problem non-trivial cobordism classes with global symmetries associated to two separate disconnected wormholes. However, we know that any sensible theory of quantum gravity should be free from global symmetries, and hence from \cite{McNamara:2019rup} any theory of quantum gravity must contain only trivial cobordism classes, meaning any $d-1$ dimensional boundary embedded inside $d$ dimensional surface must be able to transform to another $d-1$ dimensional surface by any allowed topological changing or quantum gravity operations. Therefore for the partition function to factorize especially at short distances there must exist certain topological defects which break the global symmetry associated with wormholes and makes the theory devoid of global symmetries. We will interpret these wormholes with added topological defects as our "half-wormholes". We will then calculate the spectral form factor (SFF) of the tensor model and see how gauge holonomy governs the late time behavior of SFF followed by the thermodynamics of wormhole saddles and hints of higher genus topology in the entropy spectrum. 
\\

The organization of the paper is as follows:
\begin{itemize}
\item In sections \ref{k1} and \ref{k2},  we give a brief review of the SYK model and study the behavior of the square of the partition function, first in the case when we average over the states and in the second case when we have fixed choice of coupling,  which are commonly identified as a non-averaging version of the previous one.

\item In section \ref{T1},  we explain how in the large $N$ limit the $O(N)^{q-1}$ tensor model is indistinguishable from the SYK without quench disorder.

\item In section \ref{hw},  we explicitly calculate the half-wormhole saddle points for the tensor model in the limit where dynamics of gauge holonomy can no longer be ignored. In the limit where the holonomy is identity matrix, the tensor model reduces to the regular SYK. We found that the trivial saddle points of holonomy action have wormhole contribution in the self-averaging regime whereas non-trivial saddle points have half-wormhole contributions in the non self-averaging regime.

\item In section \ref{qg},  we discuss the factorization problem of a wormhole linked Wilson operator and restate in the language of non-trivial cobordism classes or global charges associated with the disconnected wormholes.

\item In section \ref{GS},  we discuss the relationship between global symmetries and cobordism classes and why any theory of quantum gravity must be devoid of non-trivial cobordism classes.

\item In section \ref{TD},  we discuss how half-wormholes of tensor could be seen as a topological defect that breaks the global symmetry using the results that we obtain in \ref{qg}.

\item In section \ref{SFF},  we capture the late time behavior of the spectral form factor and show that plateau has $O(1)$ contributions coming from the non-trivial saddle points of holonomy.

\item In section \ref{WE},  we study the contribution from the wormhole, and half-wormhole saddles in ensemble averaged energetics,  which we have studied in terms of average energy,  average free-energy and average entropy function.

\item In section \ref{VV},  we compare our results of $O(N)^{q-1}$ with that of the SYK in the framework of without averaging over states.

\item Finally, we will conclude our results with some interesting future directions in section \ref{CC}.
\end{itemize}
\section{\textcolor{Sepia}{\textbf{ \Large Averaging vs Non-averaging}}}
\subsection{\textcolor{Sepia}{\textbf{ \large Averaged theory}}}\label{k1}
In the study of the zero-dimensional SYK model averaged over the quantity $\langle z \rangle $ vanishes therefore the most general averaged quantity $\langle z^{2} \rangle $ also defined as $z_{L}z_{R}$ is computed using the collective field formalism $"G, \Sigma "$ \cite{Jevicki:2016bwu,Polchinski:2016xgd,Gross:2016kjj,Berkooz:2016cvq,Maldacena:2016hyu,Gu:2016oyy,Saad:2018bqo,You:2016ldz,Sachdev:2010um}.  After averaging over the ensemble J with Gaussian distribution the quantity $\langle z^{2} \rangle $ is given by the following expression:
\begin{equation}
\langle z^{2} \rangle = \int d^{2N}\psi \exp \left\{\frac{N}{q}\left(\frac{1}{N}\sum_{i=1}^{N} \psi_{i}^{L}\psi_{i}^{R}\right)^{q}\right\} \label{jj}
\end{equation}
Now, after expressing \eqref{jj} in the collective field description, we get a matrix of collective fields $G_{LL}, G_{RR}, G_{LR}$ where $G_{LR}$ represents a wormhole type correlation function. By performing the integral $\Sigma $ over the imaginary axis and $G$ over the real axis \eqref{jj} is given as:
\begin{widetext}
\begin{align}
\langle z^{2} \rangle &= \frac{1}{N^N} \int_{R}dG \exp\left(\frac{N}{q}G^{q}\right)(-\partial_{G})^{N} \delta (G)= \sqrt{q} \exp\left(-\left(1 - \frac{1}{q}\right)N\right) \text{at G = 0 ~for~ large ~N}
\end{align}
\end{widetext}
By rotating the contours by:
\begin{eqnarray}
\Sigma = i\exp\left(-\frac{i\pi}{q}\right)\sigma,~~~~~G = \exp\left(\frac{i\pi}{q}\right)g,
\end{eqnarray} and doing the integration at $\phi = \frac{\pi}{q}$ over the $\sigma$ and $g$, we get non-zero value for the saddle point contribution for the variable $G_{LR}$  which is fixed by constraint $\sigma$ and can be expressed by the following equation
\begin{eqnarray}
G=\frac{1}{N}\sum_{i}\psi_{i}^{L}\psi_{i}^{R}
\end{eqnarray}
 \begin{figure}
\includegraphics[width=7cm,height=4cm]{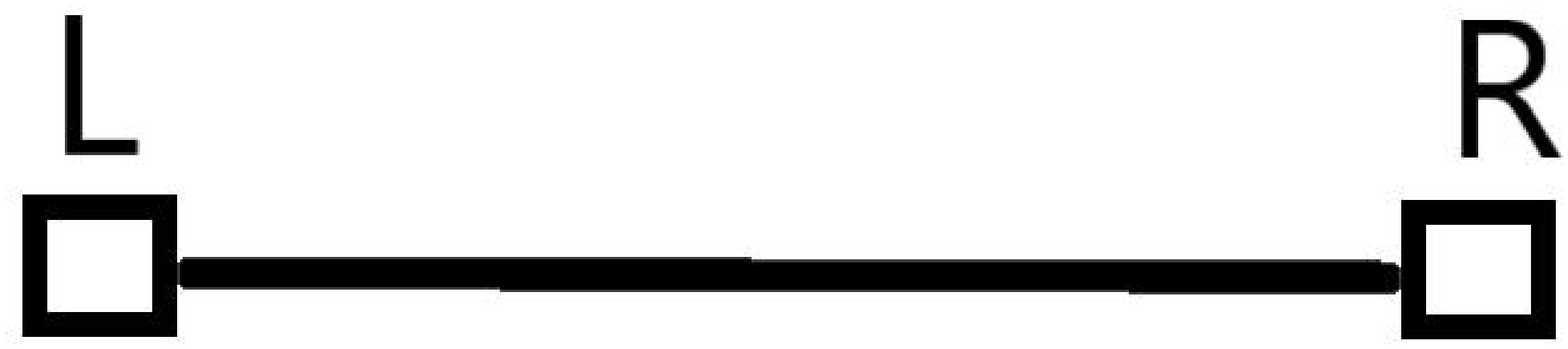}\hspace{2cm}
\caption{Correlation between left and right boundary systems.}
\label{ZLRfig}
\end{figure}
 This represents correlation between two decoupled partition functions or in euclidean gravity a wormhole connecting separate boundary regions as shown in \ref{ZLRfig}.  For more details on this aspect see the refs. \cite{Freivogel:2021ivu,Berkooz:2021ewq}.

\subsection{\textcolor{Sepia}{\textbf{ \large Non-averaged theory}}}\label{k2}
To study the behaviour of $z^{2}$ in the SYK model with fixed coupling recently in ref.\cite{Saad:2021rcu,ramsey,LR} the authors has introduced collective field variables representing the correlation between the systems $L$ and $R$ by inserting identity in the integral given by the following expression:
\begin{widetext}
\begin{align}
1 = \int dG \int \frac{d\Sigma}{2\pi i/N} \exp [-\Sigma (NG - \psi_{i}^{L}\psi_{i}^{R})] \exp \left\{\frac{N}{q}\left[G^{q} - \left(\frac{1}{N}\psi_{i}^{L}\psi_{i}^{R}\right)^{q}\right]\right\}
\end{align}
\end{widetext}
By rotating the contour as given by:
 \begin{eqnarray}
 \Sigma = i\exp\left(-\frac{i\pi}{q}\right)\sigma,~~~~~G = \exp\left(\frac{i\pi}{q}\right)g,
 \end{eqnarray} the representation of $z^{2}$ is given as:
\begin{equation}
z^{2} = \int d\sigma \Psi(\sigma)\Phi(\sigma) \label{nn}
\end{equation}
where the quantity $\Psi(\sigma)$ is given by the following expression:
\begin{equation}
\psi(\sigma) = \int \frac{dg}{2\pi /N} \exp \left[N\left(-i\sigma g - \frac{1}{q}g^{q}\right)\right]
\end{equation}
If we close look into the above equation then we can clearly observe that the above equation is peaked around $\sigma = 0$ and doesn't have any dependence on the coupling parameter at all. 

On the other hand,  the second factor in \eqref{nn} can be expressed as:
\begin{eqnarray}
\Phi(\sigma) &=&\displaystyle \int d^{2N}\psi \exp \left[i\exp\left(-\frac{i\pi}{q}\right)\sigma \psi_{i}^{L}\psi_{i}^{R} \right.\nonumber\\
&&\left.~~~~~~~~~~~~~\displaystyle+ i^{\frac{q}{2}} J_{A}(\psi_{A}^{L} + \psi_{A}^{R})- \left( i^{q}\bar{j^{2}}\psi_{i}^{L}\psi_{i}^{R}\right)^{q}\right].~~~~~\label{hl}
\end{eqnarray}
It has been shown in \cite{Saad:2021rcu} that the integral \eqref{nn} has two kinds of saddle points.  
\begin{enumerate}
\item First, near $\sigma = |1|$, here the function $\Phi = 1$ is self-averaging. These are wormhole saddles living on this region with $|\sigma| = 1$ which reproduces the exact answer for $\langle z^{2} \rangle$ for the averaged case. 

\item The second saddle point is near the region near $\sigma = 0$ where $\Phi(\sigma)$ is non self-averaging and has a weak dependence on $\sigma$ variable. The function $\Phi(\sigma)$ also has a saddle point which is exponentially peaked at $\sigma = 0$ and is described as half-wormhole in \cite{Saad:2021rcu}. 
\end{enumerate}
Therefore,  in the SYK model with fixed coupling at large $N$ the quantity $z^{2}$ could be approximately given by the following simplified expression:
\begin{equation}
z^{2} = \langle z^{2} \rangle + \Phi(0)
\end{equation}
In this paper,  we will go beyond this analysis and we will provide an  interpretation of these half-wormholes in the the context of $O(N)^{q-1}$ tensor model with fixed coupling parameter which in the large $N$ limit behaves like SYK.
\section{\textcolor{Sepia}{\textbf{ \Large Half-wormhole from $O(N)^{q-1}$ tensor Model}}}
\subsection{\textcolor{Sepia}{\textbf{ \large The $O(N)^{q-1}$ tensor Model}}}\label{T1}
\begin{figure}
\includegraphics[width=8cm,height=4cm]{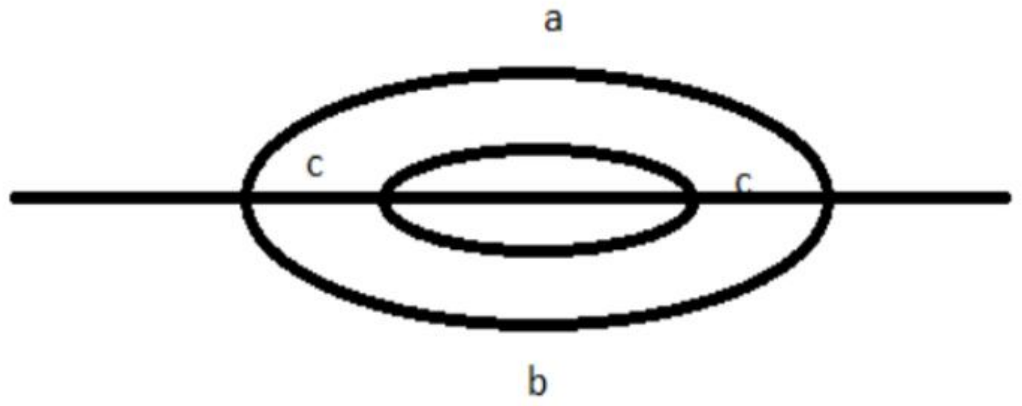}
\caption{A typical planar Feynmann graph $\mathcal{G}$ with $\omega_{\mathcal{G}} = 0$.}
\label{strands1fig}
\end{figure}

.The $O(N)^{q-1}$ tensor model is a quantum mechanical model with,  \begin{eqnarray}
q = D + 1,
\end{eqnarray}
 real fermions $\psi_{0},...\psi_{D}$.  This construction is made in such a way that for some integer $n$ each fermionic field have $n^{D}$ components which implies that the total number of fermionic field is given by \cite{Choudhury:2017tax}: 
\begin{eqnarray}
N = (D + 1)n^{D}.
\end{eqnarray}
In the large $N$ limit this model is exactly solvable as the SYK also each field $\psi_{a}$ transform according to the full symmetry group of the model which is given by:
\begin{equation} G_{0} = \prod_{a<b} G_{ab} = O(n)^{\frac{D(D+1)}{2}}, \end{equation} 
The index $a$ is the flavour index that runs from $0,1,...,D$. The action for such a model which in the large $N$ limit exhibit similar behaviour as that SYK with a fixed coupling $j$ is given by the following expression:
\begin{equation} I = \int dt \left(\frac{i}{2} \sum \psi_{i} \frac{d}{dt} \psi_{i} - i^{\frac{q}{2}} j\psi_{0}\psi_{1}...\psi_{D}\right)\label{bm}, \end{equation}
The most fundamental characteristics of the Feynman diagrams concerning this model can be found in \cite{Gurau:2010ba,Gurau:2011aq,Gurau:2011xq,Bonzom:2011zz,Bonzom:2012hw,Sasakura:1990fs}. To study the large $N$ limit of this theory it is suitable to represent each line in the diagram with $D$ strands with index $a \in \{0, 1.\dots,D\}$ labelling each vertex and the edge by $ab$ connected by vertices $a$ and $b$. These strands forms a closed loop which in turn contributes a factor of $n$. For an unordered pair $a, b \in \{0,1.\dots,D\}$ the number of closed loops $\mathcal{F}_{ab}$ is given as:
\begin{equation} \mathcal{F}=\sum_{a<b}\mathcal{F}_{ab}. \end{equation}
Here $\mathcal{F}$ is total number of faces, which will contribute a factor of the form:
 \begin{equation} n^{\mathcal{F}}=\prod_{a<b}n^{\mathcal{F}_{ab}}.\end{equation}
The large $N$ limit of this model has been studied by Witten in \cite{Witten:2016iux} given $\mathcal{J} = (a_{0}, a_{1},. \dots a_{D})$ each Feynman graph $\mathcal{G}$ can be resolved in strands of type $a_{i}b$ such that there are $D$ strands in total as shown in \ref{strands1fig}. Thus following the arguments of the matrix model each graph $\mathcal{G}$ forms a closed two manifold whose Euler characteristic $\chi_{\mathcal{J}}$ is given by:
\begin{equation} \chi_{\mathcal{J}}=2-2g_{\mathcal{J}},\end{equation} 
The model is similar to that of the SYK when the genus \begin{eqnarray}g_{\mathcal{J}} = 0. \end{eqnarray}
To see this more clearly we define the degree of the graph $\mathcal{G}$ for some non-negative genus $g_{\mathcal{J}}$ by the following expression:
\begin{equation} \omega(\mathcal{G})=\sum_{\mathcal{J}} g_{\mathcal{J}} = \sum_{\mathcal{J}}\left(1-\frac{\chi_{\mathcal{J}}}{2}\right).\end{equation}
It has been shown in ref.\cite{Witten:2016iux} that when $\omega_{\mathcal{G}} = 0$ then $g = 0$ which means that the $\mathcal{G}$ are all melonic diagrams drawn on a two-sphere. These planar diagrams are obvious planar diagrams and are constructed by taking a point at infinity in the plane. It remains to understand how $\omega_{\mathcal{G}} = 0$ produces leading-order contribution which is exactly proportional to $N$.  To visualize this more clearly,  let us define $v_{0}$ and $v_{1}$ to be the vertices and edges of the graphs which are related to each other as:
 \begin{eqnarray}
 v_{1} =\left( \frac{D + 1}{2}\right)v_{0}. \end{eqnarray}
 The numbers $v_{0}$ and $v_{1}$ do not depend on $\mathcal{J}$ however, the number of faces denoted as $v_{2, \mathcal{J}}$ does depend on the choice of $\mathcal{J}$ as:
 \begin{equation}
 v_{2, \mathcal{J}} = \sum_{i=0}^{D} \mathcal{F}_{a_{i}, a_{i + 1}}.
 \end{equation}
 From the Euler characteristics of the manifold is given by
\begin{eqnarray}
\chi_{\mathcal{J}} &=& v_{0} - v_{1} + v_{2, \mathcal{J}}\nonumber\\
& =& -\left(\frac{D - 1}{2}\right)v_{0} + \sum_{i}\mathcal{F}_{a_{i},a_{i + 1}}.\label{xx}
\end{eqnarray} 
Hence by making use of \eqref{xx} and one could finally arrive at the following simplified expression: 
\begin{equation}
\frac{2}{(D - 1)!}\omega(\mathcal{G}) = D  + \frac{D(D - 1)}{4} - \mathcal{F}
\end{equation}
Now in the large $N$ limit we take the coupling with fixed $J$ as
\begin{eqnarray}
j = \frac{J}{\displaystyle n^{\frac{D(D - 1)}{4}}}
\end{eqnarray}
Therefore a Feynman graph $\mathcal{G}$ with total number of faces $\mathcal{F}$ and vertices $v_{0}$ will contribute a factor of the following form:
\begin{eqnarray}
F(n)&=&n^{\displaystyle-\frac{D(D - 1)}{4}v_{0} + \mathcal{F}}\nonumber\\
& =& n^{\displaystyle \left(D - \frac{2}{(D - 1)!}\huge{\omega\mathcal{(G)}}\right)}\nonumber\\
& =& n^{\displaystyle D\left(1 - \frac{2}{D!}\huge{\omega\mathcal{(G)}}\right)}\nonumber\\
& =& \left(\frac{N}{D+1}\right)^{\displaystyle \left(1 - \frac{2}{D!}\huge{\omega\mathcal{(G)}}\right)}\nonumber\\
&\approx&  \left(\frac{1}{D+1}\right)^{\displaystyle \left(1 - \frac{2}{D!}\huge{\omega\mathcal{(G)}}\right)}~N~~~~{\rm at ~large~N}\label{yy},~~~
\end{eqnarray}
for all $\omega\mathcal{(G)} \geq 0 $ the leading order contribution in the large $N$ limit are at most proportional to $N$.  However precisely at $\omega\mathcal{(G)} = 0$ the leading order contribution is exactly proportional to $N$ i.e 
\begin{equation}
F(n)=n^D=\left(\frac{N}{D+1}\right) \ \text{when} \  \omega(\mathcal{G}) = 0\label{hh}
\end{equation} 
Here we have used the relationship between $n$ and $N$ explicitly for both cases.
Thus a graph with $g_{\mathcal{J}} = 0$ is planar and can be drawn on a two-sphere by only one way.  However, if $\omega(\mathcal{G})$ is also zero then it generates the leading order diagrams of the SYK model that can be drawn on a two-sphere in many possible ways.

Though in the large $N$ limit the quantum mechanical model seems to agree with the SYK,  there exist new dynamics at sub-leading order $\displaystyle \frac{1}{N}$ in the form of new light modes due to the time-dependent $O(N)^{q-1}$ transformations, there also exist modes in the original the SYK that arise due to conformal diffeomorphism.  Hence the total number of new light modes is given by $\displaystyle (q-1)\frac{N^{2}}{2}$ which is quantitatively very large in the large $N$ limit and whose dynamics is governed by effective sigma action as shown in the ref.~\cite{Choudhury:2017tax}.  These new light modes have an interesting bulk interpretation as they might represent gauge fields propagating in the $AdS_{2}$ background.

Most significantly,  in this tensor model which we have described in \eqref{bm} also has holonomy contributions in addition to matter fields whose effective action takes a simple universal form $S_{eff}(U)$.  In sec \ref{hw},  we will show how the dynamics of gauge holonomy contributes in restoring the factorization of the partition function and unitarizing black hole dynamics by giving rise to half-wormhole saddle points, it is when the holonomy of the gauge group is identity matrix the quantum mechanical model is indistinguishable from the original SYK.

 \subsection{\textcolor{Sepia}{\textbf{ \large Computation of half-wormhole saddle points}}}\label{hw}
In this section, we will compute the quantity $z_{LR}$ in the collective field description which unlike $SYK$ transform under the gauge group $O(N)^{q-1}$ melonic theory.  We will follow the same procedure as mentioned in ref \cite{Saad:2021rcu} since the quantum mechanical tensor model we will be working with has a fixed coupling chosen out of an Gaussian Unitary Ensemble (GUE). We introduce an identity element in the partition function by the following fashion:
\begin{widetext}
  \begin{align}
  1 = \int dG \int \frac{d\Sigma}{2\pi i/N} \exp [-\Sigma_{b}^{a} (NG_{b}^{a} - (\psi_{a})^{L}(\psi_{b})^{R}]\exp \left\{\frac{N}{q}\left[(G^{a}_{b})^{q} - \left(\frac{1}{N}(\psi_{a})^{L}(\psi_{b})^{R}\right)^{q}\right]\right\}
  \end{align}
  \end{widetext}
  where the delta function is represented as an integral over $\Sigma$. The partition function now becomes
  \begin{widetext}
  \begin{align}
  Z_{LR} = \int d^{2N}\psi \exp \left \{i^{\frac{q}{2}} \sum J((\psi_{a})^{L} + (\psi_{b})^{R})\right \}\int_{R} dG~ \delta\left(G_{b}^{a}  - \frac{1}{N}(\psi_{a})^{L}(\psi_{b})^{R}\right)\exp \left\{\frac{N}{q}\left[(G_{b}^{a})^{q} - \left(\frac{1}{N}(\psi_{a})^{L}(\psi_{b})^{R})\right)^{q}\right]\right\}
  \end{align}
  \end{widetext}
  We then rotate the integration contour of $G$ and $\Sigma$ with the amount $\displaystyle \phi = \frac{\pi}{q}$ as described in the previous section for $q$ number of Majorana fermions.  In the original SYK model, this procedure splits the integral into two parts,  (1) one described by $\psi(\sigma)$ and (2) the other by $\phi(\sigma)$.  However, in the quantum mechanical model to calculate the expression for $Z_{LR}$ one has to take into account the holonomic degrees of freedom due to the gauge group $O(N)^{q-1}$,  which are absent in the original SYK model without having any holonomy contribution.  In the large $N$ limit, the saddle points of the SYK action in the collective field description effectively captures the free energy of the system. Therefore at fixed temperatures when the holonomy of the gauge group is identity the saddle points of the tensor model coincides with that of the SYK model,  which is quite expected from this analysis.
  
Therefore,  when the holonomy of the gauge group is identity the saddle points of $Z_{LR}$ is effectively given by \eqref{hl} when we average over the coupling of original SYK i.e.
  \begin{eqnarray}
  \langle \Phi(\sigma) \rangle &=& \int d^{2N} \psi \exp \left\{i\exp\left(-\frac{i\pi}{q}\right)\sigma\psi_{i}^{L}\psi_{i}^{R}\right\} \nonumber\\
  &=& \left[i\exp\left(-\frac{i\pi}{q}\right)\sigma\right]^{N}.
  \end{eqnarray}

  However,  in the limit when each $U_{i}$ which is an $O(N)$ matrix that represents the holonomy in $i^{th}$ factor in the gauge group $O(N)^{q-1}$)is not the identity matrix i.e.  when the dynamics of the gauge group $V(t)$ becomes effective,  where $V(t)$ is an arbitrary group element of $O(N)^{q-1}$, the two decoupled boundary correlator is non-zero as follows:
   \begin{eqnarray} G_{LR} \not = \frac{1}{N}\psi_{i}^{L}\psi_{i}^{R}.\end{eqnarray}
Unlike when the holonomy of the gauge group is identity,  this can be seen as follows,  since the holonomy matrices $U_{m}$ are unitary its eigenvalues can be expressed in the form $\displaystyle\exp(i\theta^{n}_{m})$,  where $n$ runs from $1$ to $N$.  Consequently,  the corresponding eigenvalue density function could be defined as
  \begin{equation}
  \rho_{m}(\theta) = \frac{1}{N}\sum_{n=1}^{N}\delta(\theta - \theta_{m}^{n}).
  \end{equation}
  Since the eigenvalues can be written as an integral the above mentioned Dirac delta function at large $N$ can be represented by the following expression:
  \begin{equation}
  \frac{TrU_{m}^{n}}{N} = \frac{\displaystyle \sum_{j=1}^{N} \exp(in\theta_{m}^{j})}{N} = \int \rho_{i}(\theta) \exp(in\theta) = \rho_{m}^{n}
  \end{equation}
  where it is important to note that,  $m$ runs from $1$ to $q-1$ and $n$ runs from $0$ to $\infty$ which the eigenvalue distribution in the Fourier mode. Following from ref.  \cite{Choudhury:2017tax} in the large $N$ limit, the integral over the eigenvalues $\theta_{m}^{n}$ can be morphed into a path integral over the eigenvalue density function as given by the following expression:
  \begin{widetext}
  \begin{eqnarray}
  Z(x) = \int \prod_{i=1}^{q-1} \mathcal{ D}\rho ~\exp \left[ \frac{1}{2} \sum_{n=1}^{\infty} \left(-N^{2}\sum_{m=1}^{q-1}\frac{|\rho_{m}^{n}|^{2}}{n} - 2N_{F}N^{q-1}(-x)^{n}\frac{ \left(\displaystyle\prod_{m=1}^{q-1}\rho_{m}^{n}\right)}{n}\right)\right]\label{zk}
  \end{eqnarray}
  \end{widetext}
The above equation has two kinds of terms: 
\begin{itemize}
\item First kind which is proportional to $N^{q-1}$,

\item The second kind which is proportional to $N^{2}$.
\end{itemize}
Here $x = e^{-\beta m}$, where $m$ is the mass of the fermions which is taken to be positive and which we suppose arises due to the interaction of the gauge fields when the holonomy is switched on. when $x$ is of unit order the partition function reduces to $\Phi(\sigma)$. In other words when holonomy of the gauge group is identity matrix the above function reduces to the partition function of the original SYK model.  However, when $\displaystyle x = \frac{\alpha}{pN^{q-3}}$ where $p = N_{F}$ the contributions coming from both dynamics of gauge and energy compete with each other thus in the limit where $N \rightarrow \infty$ and $\alpha$ is held fixed the partition function as stated in \eqref{zk} simplifies to the following form:
\begin{equation}
Z(\alpha) = \int \prod_{i=1}^{q-1}dU_{i}\exp(-S_{eff}(U_{i}))\label{jo}
\end{equation} 
where the effective action $S_{eff}(U_{i})$ is given by:
\begin{equation}
S_{eff} = -\frac{\alpha}{N^{q-1}}\left(\prod_{i=1}^{q-1} TrU_{i}\right)
\end{equation}
The above integral \eqref{jo} is related to integral over unitary matrices and can be solved easily in the large $N$ limit where the integral is taken over the $U_{1}$ since $Tr U_{2},....Tr U_{q-1}$ are all constant,  which reduces \eqref{jo} to the following simplified form:
\begin{equation}
Z_{SU} = \int dU_{1}\exp\left(\frac{N}{g_{1}}(Tr U_{1} + Tr U_{1}^{\dagger})\right)\label{kj},
\end{equation}
where we have:
\begin{equation}
\frac{1}{g_{1}} = \alpha\rho_{2}^{1}\rho_{3}^{1}.....\rho_{q-1}^{1} = \alpha u_{2}u_{3}....u_{q-1}.
\end{equation} 
Here we refer $\rho_{m}^{1} = u_{m}$.  The saddle points that extremizes the equation \eqref{kj},  which has been calculated in ref. \cite{Choudhury:2017tax} as:
\begin{equation}
 u = \large \left\{
	\begin{array}{ll}
		 \displaystyle \alpha u^{q-2} & \mbox{if } u \leq \frac{1}{2} \\ 
		 \displaystyle 1- \dfrac{1}{4 \alpha u^{q-2}}, & \mbox{if } u > \frac{1}{2}.
	\end{array}
\right.
\end{equation}
since all of $u_{m}$ and $g_{m}$ are equal we define $u$ as the common saddle point value of $u_{m}$.  Once the above mentioned saddle point solution is determined then one can immediately write down the expression for the partition function as stated in \ref{jo} in the large $N$ as \cite{Choudhury:2017tax}:
\begin{equation} \label{par}
Z(\alpha)=\exp\left(-\frac{N^2}{2}V(u,\alpha)\right),
\end{equation}
where the holonomy dependent effective potential $V(u,\alpha)$ in this context can be computed as \cite{Choudhury:2017tax}:
\begin{equation}
V(u,\alpha)=(q-1)f(u)-2\alpha~ u^{q-1},
\end{equation}
where we define the holonomy dependent new function $f(u)$ by the following expression \cite{Choudhury:2017tax}:
\begin{equation}
 f(u) = \large \left\{
	\begin{array}{ll}
	 \displaystyle 0 & \mbox{if } u = 0 \\ 
		 \displaystyle u^{2} & \mbox{if } u \leq \frac{1}{2} \\ 
		 \displaystyle \frac{1}{4}- \frac{1}{2}\ln\left[2(1-u)\right], & \mbox{if } u > \frac{1}{2}.
	\end{array}
\right.
\end{equation}
After substituting back all these expression in the expression \ref{par} and after doing a little bit of algebraic manipulation we get the following simplified expression for the holonomy dependent partition function at the large $N$ limit,  which is given by:
\begin{widetext}
\begin{equation}\label{Sayang}
 Z(\alpha) = \large \left\{
	\begin{array}{ll}
	 \displaystyle 1 & \mbox{if } u= 0 \\ \\
		 \displaystyle \exp\left(-\frac{N^2}{2}\left[(q-1)u^2-2\alpha~u^{q-1}\right]\right) & \mbox{if } u \leq \frac{1}{2} \\ \\
		 \displaystyle 2^{\displaystyle \frac{N^2}{4}(q-1)}~\exp\left(-\frac{N^2}{8}(q-1)\right)~\left(1-u\right)^{\displaystyle \frac{N^2}{4}(q-1)}~\exp\left(N^2\alpha~ u^{q-1}\right), & \mbox{if } u > \frac{1}{2}.
	\end{array}
\right.
\end{equation}
\end{widetext}
The presence of these sub-dominant saddle plays an important role in unitary evolution of correlators around black hole. Thus we see that when the holonomy of the gauge group is identity the saddle points of $Z_{LR}$ is effectively given by $\langle \Phi(\sigma) \rangle$ however for fluctuations around the identity matrix where the dynamics of the holonomy are non-trivial we see that  the identity from the ref.\cite{Saad:2021rcu} also persists for quantum mechanical model as stated in \eqref{bm} which is:
\begin{widetext}
\begin{eqnarray}
Z_{LR} = ({\rm wormhole~ saddle~ at}~ u = 0 ) + ({\rm "half-wormhole"~ saddle~ with}~ |u| = 1)
\end{eqnarray}
\end{widetext}
We see that at $u=0$ the partition function has value $Z(\alpha)=1$ which means precisely at $u=0$,  $Z(\alpha) = \langle \Phi(\sigma) \rangle=1$ and therefore wormhole saddle turns out to be in the self-averaging regime.  However,  for $|u|$ around $1$ the value of the partition function will be given by the second and the third line contributions appearing in the above equation \eqref{Sayang},  which are approximately the contributions from wormhole saddles. 

Further,  one can use equation \eqref{Sayang} for the holonomy dependent partition function at the large $N$ to understand how exactly in the wormhole saddle at $u=0$ and in the half-wormhole saddle around $|u|=1$ the averaged energy as well as the free energy contribute from this particular $O(N)^{q-1}$ tensor model computation.  Doing further a bit of simple algebra the averaged energy as well as the free energy from the mentioned system at different saddle point contributions can be explicitly computed as:
\begin{widetext}
\begin{equation}\label{Sayan2}
 E(\alpha) =\alpha \partial_{\alpha}\ln Z(\alpha)=\large \left\{
	\begin{array}{ll}
	 \displaystyle 0~~~~~~~~~~~~~~~ & \mbox{if } u= 0 \\ \\
		 \displaystyle N^2\alpha~u^{q-1} & \mbox{if } u \leq \frac{1}{2} \\ \\
		 \displaystyle N^2\alpha~u^{q-1}, & \mbox{if } u > \frac{1}{2}.
	\end{array}
\right.
\end{equation}
and
\begin{equation}\label{Sayan3}
 F(\alpha) =-\frac{1}{\alpha} \ln Z(\alpha)=\large \left\{
	\begin{array}{ll}
	 \displaystyle 0~~~~~~~~~~~~~~~ & \mbox{if } u= 0 \\ \\
		 \displaystyle \frac{N^2}{2}\left[\frac{1}{\alpha}(q-1)u^2-2~u^{q-1}\right] & \mbox{if } u \leq \frac{1}{2} \\ \\
		 \displaystyle N^2\Bigg\{u^{q-1}+\frac{1}{4\alpha}(q-1)\bigg[\frac{1}{2}-\ln 2-\ln(1-u)\bigg]\Bigg\} & \mbox{if } u > \frac{1}{2}.
	\end{array}
\right.
\end{equation}
\end{widetext}
Last but not the least,  one can push forward this analysis to compute the contribution from the entropy function at the large $N$ to understand how exactly in the wormhole saddle at $u=0$ and in the half-wormhole saddle around $|u|=1$ contribute in the present context,  which gives us finally:
\begin{widetext}
\begin{equation}\label{Sayan4}
 S(\alpha) =\left(\ln Z(\alpha)-E(\alpha)\ln\frac{\alpha}{N^{q-3}p}\right)=\large \left\{
	\begin{array}{ll}
	 \displaystyle 0 & \mbox{if } u= 0 \\ \\
		 \displaystyle \frac{N^2}{2}\left[(1-q)u^2+2\alpha~u^{q-1}\right]-N^2\alpha~u^{q-1}\ln\frac{\alpha}{N^{q-3}p} & \mbox{if } u \leq \frac{1}{2} \\ \\
		 \displaystyle N^2\Bigg\{-\alpha u^{q-1}+\frac{1}{4}(1-q)\bigg[\frac{1}{2}-\ln 2-\ln(1-u)\bigg]\Bigg\}\\
		 \displaystyle ~~~~~~~~~~~~~~~~~~~~~~~~~ -N^2\alpha~u^{q-1}\ln\frac{\alpha}{N^{q-3}p}  & \mbox{if } u > \frac{1}{2}.
	\end{array}
\right.
\end{equation}
\end{widetext}
\subsection{\textcolor{Sepia}{\textbf{ \large The Factorization problem}}}\label{qg}
\begin{figure}
\includegraphics[width=9cm,height=8cm]{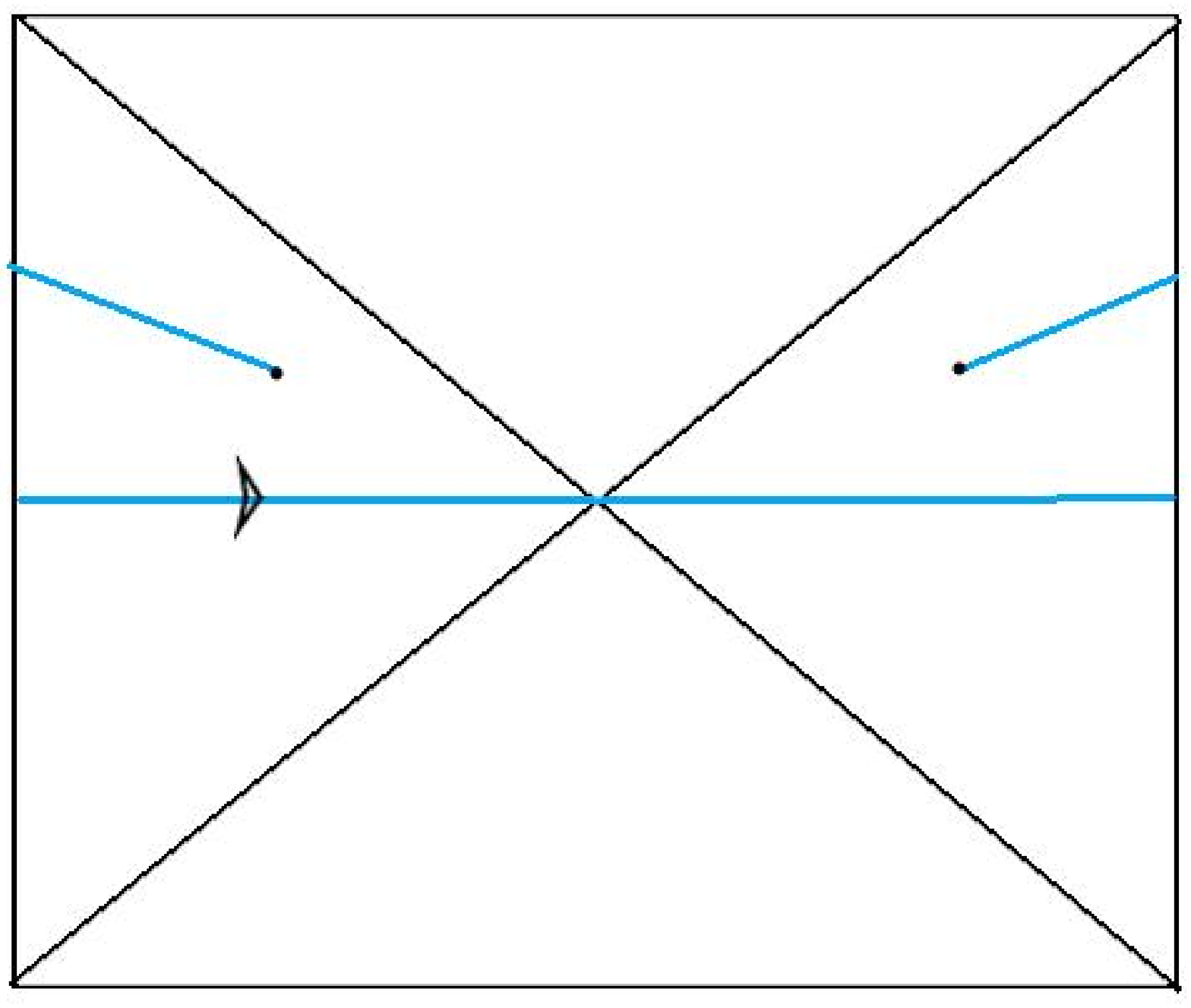}\hspace{3cm}
\caption{The blue line represent wormhole threaded Wilson line on the AdS and AdS-Schwarzschild background. The black dot represent local operators creating positive and negative charges.}\label{wilson2fig}
\end{figure}

We will now focus on the problem of factorization of gauge fields in the bulk in terms of boundary operators in the dual CFT.  We will consider a wormhole threading Wilson line that runs through the bulk connecting the two boundaries since the property of Wilson line makes it necessary for the two boundaries to be connected through the bulk as shown in fig \ref{wilson2fig}. However,  such a wormhole threaded Wilson line couldn't be reconstructed as a CFT operators as cutting the Wilson lines into two parts gives us operators that are not individually gauge invariant \cite{Gurau:2016cjo,Harlow:2015lma,Hamilton:2006az,Kabat:2011rz,Kabat:2013wga,ss}:
\begin{widetext}
\begin{equation}
W_{n} = \exp\left(i~n \int_{L}^{R} A\right) = \exp\left(i~n \int_{L}^{0} A\right)~\exp\left(i~n \int_{0}^{R} A\right)~~~\forall  (n \in Z)
\end{equation}
\end{widetext} 
This is known as the well-known \textit{factorization problem} since in the Hilbert space of the theory any two CFT operators can be factorized as the sum of tensor product of one-CFT operators which are themselves gauge invariant under the bulk gauge transformations. Thus it has been shown that for the two CFT Hilbert space to factorize in the presence of wormhole-threading Wilson lines the CFT must possess local operators that are charged under the symmetry current dual to the gauge field because if we calculate the expectation value of the  operator $Q_{L} - Q_{R}$ on the thermofield double with a Wilson line we get states with non-trivial single CFT charged operator $Q$.  Now cutting the Wilson line in the fundamental representation with the pair of opposite charges at the ends by the following fashion: 
\begin{equation}
W' = \overleftarrow{\phi_{l}}^{\dagger}\overrightarrow{\phi}_{l+1}
\end{equation} 
we can then factorize this newly defined operators in the left (L) and right (R) half of the CFT. 

\begin{figure}
\includegraphics[width=9cm,height=5cm]{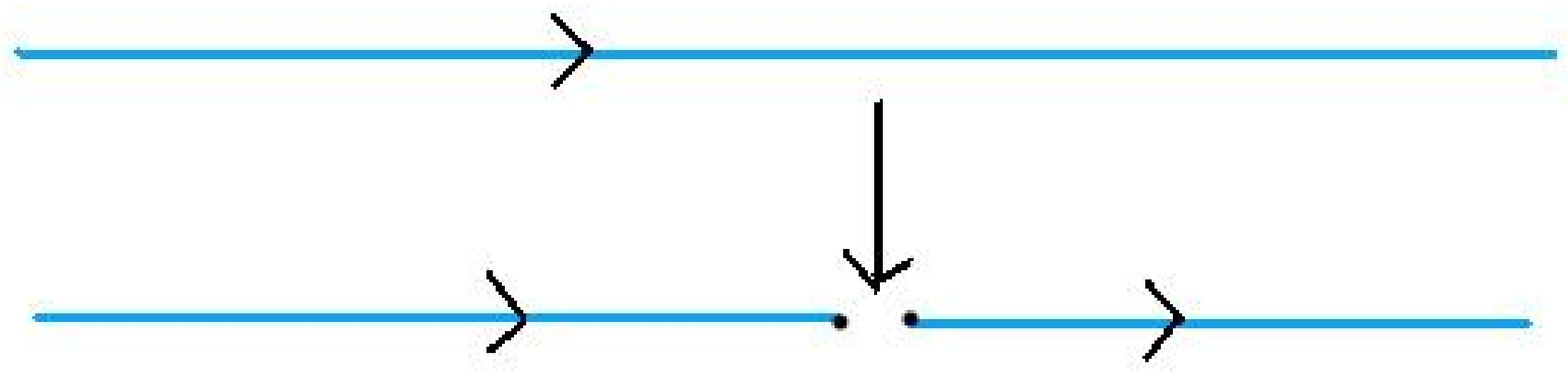}\hspace{2cm}
\caption{The two operators are proportional to each other under the Renormalization group flow. The spilt Wilson line with charges at opposite ends factorizes in the left and right half CFT.}\label{RG2fig}
\end{figure}

 One could argue that the now defined operator in the fundamental charged representation is not the one we started with however, these operators are proportional to each other in the low-energy correlation function under the renormalization group (RG) flow as shown in fig \ref{RG2fig}.  However, in short distances, this mechanism would only resolve the factorization problem if the gauge field comes out to be emergent right before the Planck scale since any observer at such short distance will notice the difference between Wilson operators connecting the bulk.  In the next section,  we will show that how the above defined factorization could be defined in terms of non-trivial cobordism classes.

\section{\textcolor{Sepia}{\textbf{ \Large Global symmetries and Cobordism classes}}}\label{GS}
  Cobordism refers to an equivalence relation more general than diffeomorphism between two smooth manifolds. In other words, when two manifolds are related to each other by an allowed set of topological operations then they are said to be cobordant and are said to be equivalent to each other. Now we know that any theory of quantum gravity with two $n$-dimensional topologies namely $M, N$ are cobordant i.e $M \sim N$ if they are related by a sequence of topological operations, and one should allow only those topological changing processes which are dynamically allowed in our theory of quantum gravity \cite{McNamara:2019rup,Gaiotto:2014kfa}. Thus the cobordism groups are defined as
  \begin{equation}
\Omega_{k}^{QG}  = \{ \text{Compact, Closed n-dimensional backgrounds} \} / \sim
  \end{equation} 
It has been shown in \cite{McNamara:2019rup} that the presence of a non-trivial cobordism group $\Omega_{k}^{QG}$ means there exists a global symmetry with charges represented by classes,
\begin{equation}
[M] = \Omega_{k}^{QG}
\end{equation}
We know that any theory of quantum gravity should be free from global symmetries other it represents an inconsistency of the following $d$-dimensional theory which implies that in the context of AdS/CFT correspondence the cobordism group must be trivial i.e $\Omega_{k}^{QG} = 0$. In case if we have an approximate consistent theory of quantum gravity with non-trivial cobordism classes then we need to add topological defects in order to cancel the charge to make sense of the full theory \cite{Harlow:2018tng,Banks:2010zn,Distler:2009ri}. In other words there exist topological defects such as singularities which breaks the global symmetry and produce new cobordism classes which were not previously connected. The map
\begin{equation}
\Omega_{k}^{QG} \rightarrow \Omega_{k}^{QG + \text{defects}}
\end{equation}
represents the transition of approximate theory to full theory with no global symmetry.
\section{\textcolor{Sepia}{\textbf{ \Large Half-wormholes as topological defects}}}\label{TD}
\begin{figure}
\includegraphics[width=9cm,height=8cm]{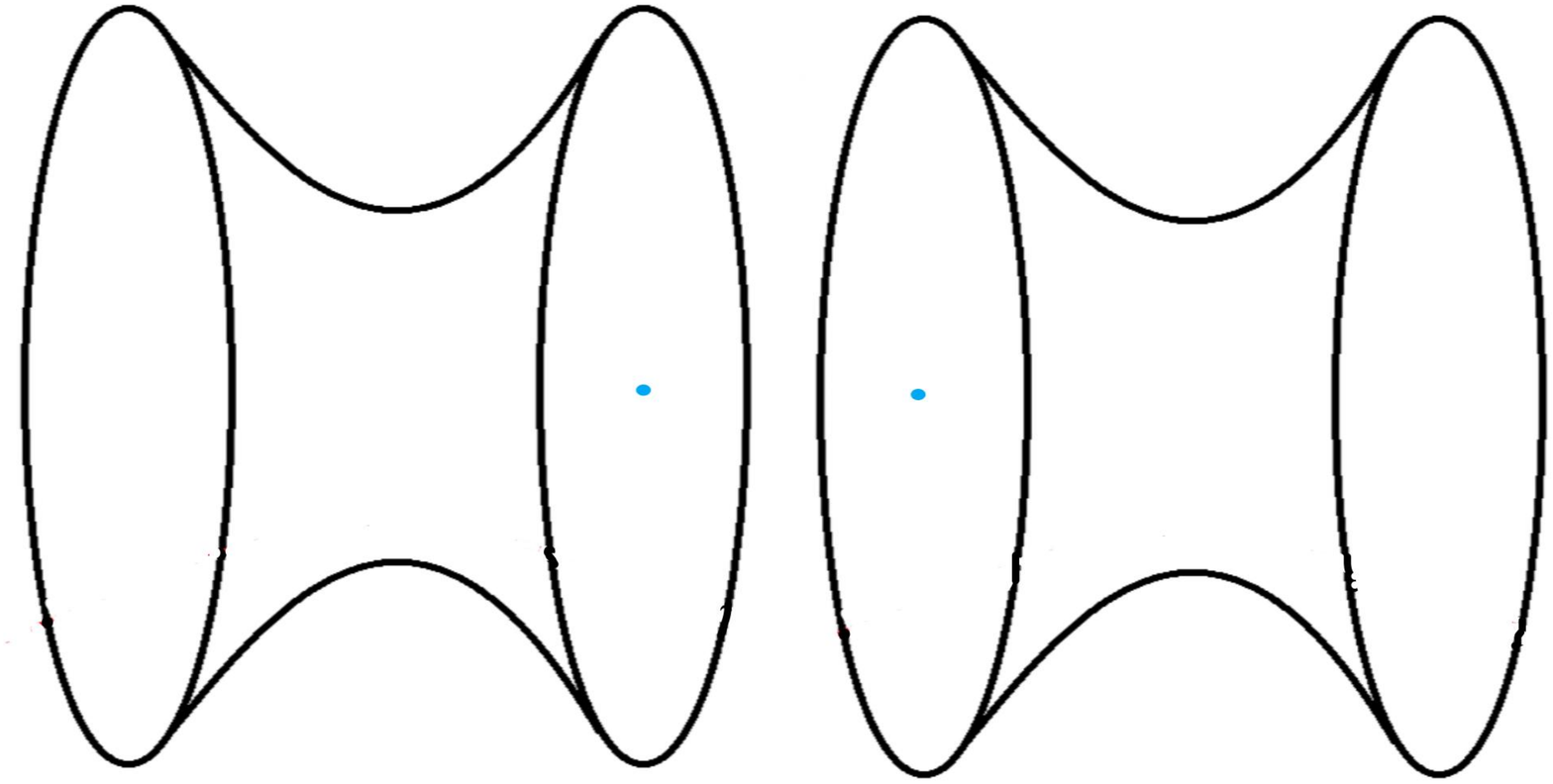}\hspace{3cm}
\caption{The spilt Wilson line with charge field at opposite ends at short distances could be represented as a pair of disconnected wormhole with non-trivial cobordism classes.}\label{Diskfig}
\end{figure}

\begin{figure}
\includegraphics[width=11cm,height=11cm]{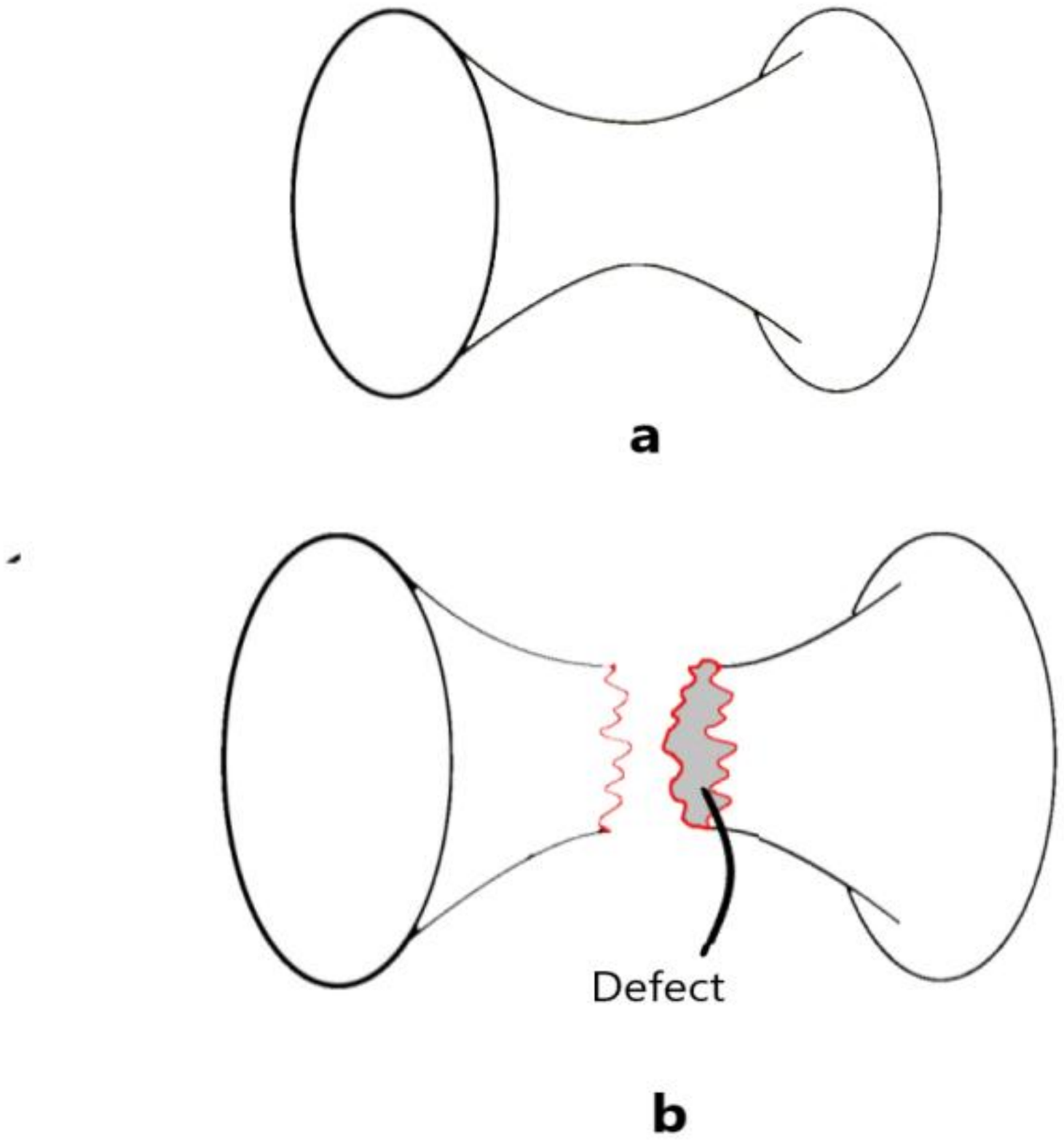}\hspace{3cm}
\caption{The figure (a) represents the wormhole in the limit where the Wilson operator $W$ and $W'$ are indistinguishable. The inability of the Wilson operator $W$ to factorize at short distances could be restated in the language of non-trivial cobordism classes which prohibits the boundaries of wormhole in \ref{Diskfig} to merge. Figure (b) is a wormhole with topological defect represented by red wavy lines which introduces new cobordism classes in order for the boundaries to connect and breaks the global symmetry.  }\label{half3fig}
\end{figure}

 In this section we will restate the factorization problem linked with wormhole threaded Wilson lines in terms of global charges associated with the wormhole at short distances. From sec \ref{qg} we recall that the operator $W$ is proportional to $W'$ only at low energies and thus would fail to factorize at short distances. So instead of taking a single wormhole threaded Wilson line let us take two wormholes with end charges of operator $W'$ resting on the end and mouth of two disconnected wormholes as shown in fig \ref{Diskfig}. At low energies, the difference between the two figures tends to vanish and we are again back to the configuration where the operator with fundamental charges seems to factorize. However, at short distances purely from the gravitational perspective, the inclusion of two wormholes seems to create a problem for the partition function to factorize. 
One could now restate the problem as follows, we now from sec \ref{qg} that at short distances any observer would be able to figure out the difference between operator $W$ and $W'$ thus in the context of wormholes as shown in fig \ref{Diskfig} the problem of factorization seems to get converted to the problem of non-trivial cobordism classes. To see this more clearly let us calculate the product one point charged correlators:
\begin{equation}
\langle O_{q} \rangle_{\beta}\langle O^{\dagger}_{q} \rangle_{\beta} = 0\label{hfhfd}~~~~~~(\rm for ~fig~ \ref{half3fig}(a))
\end{equation}
and
\begin{equation}
\langle O_{q} \rangle_{\beta}\langle O^{\dagger}_{q} \rangle_{\beta} = \exp(-S)\label{djj}~~~~~~(\rm for ~fig~ \ref{half3fig}(b))
\end{equation}
Whether the above correlation function yields a non-zero value depends on under what kind of symmetries the bulk field transforms if the symmetry is gauged then the correlation function vanishes as shown in \eqref{hfhfd} but if the symmetry is global symmetry then the contribution of the wormhole will yield a non-zero answer as in \eqref{djj} where $S$ is the entropy of the system. \cite{Belin:2020jxr,DAlessio:2016rwt,Balasubramanian:2011ur,Sonner:2017hxc,Anous:2019yku,Garrison:2015lva,Susskind:2021omt}. Since at a short distance the Wilson operator $W'$ no longer seems to factorize since the operator $W'$ is not equal to $W$ at high energies, this means that the operator is no longer gauge invariant therefore the boundary to boundary correlation function will be non-zero.  Now in this case two things could happen either the bulk gauge field must be emergent as shown in \cite{Harlow:2015lma} or since we know that wormhole seems to carry information about the variance of observables as,
\begin{equation}
\langle \phi \phi \rangle_{\rm wormhole} = \bar{|\langle O \rangle|^{2}}
\end{equation} 
Therefore equivalently one could attach global charges to the object in fig \ref{Diskfig} which in the language of cobordism means that the boundaries of two wormholes don't seem to be connected in the bulk by any sequence of allowed topological changes or quantum gravity operations. Now, this seems to be a much bigger problem as it will allow wormholes to have global charges transforming under global symmetries which can't happen in any theory of quantum gravity. Therefore in order to cancel these global charges there should exist certain topological defects as in fig \ref{half3fig} which would break the global symmetry and construct new cobordism classes which will restore the factorization problem even at short distances. We call these wormholes with added a topological defects our half-wormholes whose contribution in the SYK model with fixed coupling also restores the factorization into the left and right parts of the CFT operator.
\par{From the above discussion we like propose that the duality between wormholes and half-wormholes could be seen as an electromagnetic duality particularly in the limit of topological field theory where electro-magnetic duality in $4D$ is analogous to geometric Langlands correspondence \cite{Kapustin:2006pk}. This could be seen as follows, from the above discussions we know that the problem of factorization of Wilson operator can be restated in terms of wormholes having global charges or non-trivial cobordism classes. Therefore in order to break the global symmetry, one needs to add certain topological defects to the wormhole, particularly at short distances. In case the defect is a kind of singularity it is then given by the holomorphic $G$ bundles on $P^{1}$defined as $Bun_{G}P_{1}$ which then could be translated in the form of an irreducible representation of Langlands dual group as $Irr(G')$. Given a representation $V$ of Langlands dual group we get a t'hooft operator labeled by $H_{V}$ thus the operator $W'$ at short distances is now represented by representations of the Langlands dual group. Now in deriving Langlands correspondence one exchanges electric and magnetic observables i.e Wilson and t'hooft loop which in our case are wormholes and half-wormholes.} 
\section{\textcolor{Sepia}{\textbf{ \Large Spectral form factor}}}\label{SFF}
In this Section, we will compute the two point spectral form factor for our quantum mechanical model \eqref{bm}. We will see how the presence of non-trivial gauge holonomy effectively describes the ramp behaviour for fixed choice of coupling. For a statistical ensemble (GUE) governed by random unitary matrix \cite{Gaikwad:2017odv,Choudhury:2018lcb,Choudhury:2018rjl,Grassi:2019txd,Guhr:1997ve,Cotler:2017jue,Kodama:2008zra} we define our thermofield double state as:
 \begin{equation}\label{HTFD}
| \Psi \rangle_{\bf TFD} = \frac{1}{\sqrt{|L|}}\sum_{n,m = 1}^{N} U_{n,m}|m\rangle_{L}|n\rangle_{R}
 \end{equation}
where $L$ is the value of $Z(\beta + it)$ at $t=0$, which is basically $Z(\beta)$ expressed in \eqref{jo}.  We will study the normalized version of the quantity $ \langle Z(\beta + it)Z(\beta - it)\rangle$, which is commonly identified to be the Spectral Form Factor and from the present analysis we will try to observe how it differs from the regular SYK model computation.  Now to proceed with this computation we take disordered average over the ensemble,  especially in the presence of holonomy whose density matrix will be constructed and can be further used for the computation of the spectral form factor,  which is the present context turns out to be:
\begin{widetext}
\begin{eqnarray}
  {\bf SFF}= \frac{\langle Z(\beta + it)Z(\beta - it)\rangle}{\langle |Z(\beta)|^{2}\rangle} &=& \frac{{}_{\bf TFD} \langle \Psi | |Z(\beta + it)Z(\beta - it) | \Psi \rangle_{\bf TFD} }{{}_{\bf TFD} \langle \Psi |Z(\beta)|^{2}| \Psi \rangle_{\bf TFD}} \nonumber\\
  &=&\sum^{N}_{n,n',m,m'=1}\int dU \langle \rho^{n}_{m} \rho^{n'}_{m'} \rangle \nonumber\\
  &=& \frac{1}{|L|N^2}\sum^{N}_{n,n',m,m'=1}\int d\theta~\delta(\theta - \theta^{n}_{m})\delta(\theta - \theta^{n'}_{m'}) + \frac{1}{N^2}\sum^{N}_{n,n',m,m'=1}\int dU\langle \rho^{n}_{m} \rangle \langle \rho^{n'}_{m'}
  \rangle\nonumber\\
  &=& \frac{1}{|L|N^2}\sum^{N}_{n,n',m,m'=1}\delta(\theta^{n}_{m} - \theta^{n'}_{m'}) + \frac{1}{N^2}\sum^{N}_{n,n',m,m'=1}\int dU\langle \rho^{n}_{m} \rangle \langle \rho^{n'}_{m'} \rangle\label{qq}
\end{eqnarray}
\end{widetext} 
 Since we have explicitly shown the presence of half-wormhole saddle points in the presence of non-trivial gauge holonomy where non-trivial saddle points, we suspect the partition function to effectively factorize, therefore \eqref{qq} which describes the non-trivial saddle points can be written as: 
 \begin{widetext}
 \begin{eqnarray}
 \int dU\langle \rho^{n}_{m} \rangle \langle \rho^{n}_{m}\rangle&=& \int dU \langle Tr  U_{1L} \rangle \langle Tr U^{*}_{1R} \rangle \nonumber\\
 &=&\underbrace{\frac{1}{|L|^{2}N^2}\sum^{N}_{n,n',m,m'=1} \delta_{n,n'}^{L}\delta_{m,m'}^{R}}_{\textcolor{red}{\rm Contribution~from ~wormhole~saddle~at~u=0}}\nonumber\\
 &&~~~~~+\underbrace{\underbrace{\frac{1}{|L|^{2}N^2}\sum^{N}_{n,n',m,m'=1} \delta_{n,n'}^{L}\delta_{m,m'}^{R}~\mathcal{J}_{1} (N,q,\beta)}_{\textcolor{red}{\rm Contribution~from ~half-wormhole~saddle~at~u\leq \frac{1}{2}}}+\underbrace{\frac{1}{|L|^{2}N^2}\sum^{N}_{n,n',m,m'=1}  \delta_{n,n'}^{L}\delta_{m,m'}^{R}~\mathcal{J}_{2} (N,q,\beta)}_{\textcolor{red}{\rm Contribution~from ~half-wormhole~saddle~at~u>\frac{1}{2}}}}_{\textcolor{blue}{\rm Contribution~from ~half-wormhole~saddle~with~|u|=1}}.~~
 \end{eqnarray}
 \end{widetext}
In the above equation is an integral over Haar measure ensemble average over unitary matrices which can be easily evaluated in the present context.  See ref. \cite{Cotler:2016fpe} more discussion on this topic.  Here we have introduced two integrals $\mathcal{J}_{1} (N,q,\beta)$ and $\mathcal{J}_{1} (N,q,\beta)$ over the holonomy,  which are given by:
\begin{widetext}
\begin{eqnarray}
 \mathcal{J}_{1} (N,q,\beta):&=& \int_{u=\Delta > 0}^{1/2} du~ u^{2}~ \exp\left(-{N^2}\left[(q-1)u^2-4\beta~u^{q-1}\right]\right),\\
 \mathcal{J}_{2} (N,q,\beta):&=& 2^{\displaystyle \frac{N^2}{2}(q-1)}~\exp\left(-\frac{N^2}{4}(q-1)\right)\int_{u=1/2 + \Lambda(>0)}^{1} du~ u^{2}~\left(1-u\right)^{\displaystyle \frac{N^2}{2}(q-1)}~\exp\left(N^2 2\beta~ u^{q-1}\right).
\end{eqnarray}
\end{widetext}
where physically $\Delta$ represents small positive shift from $u=0$ and $\Lambda$ represent small positive shift from $u=\frac{1}{2}$ saddle points.

These integrals further cannot be exactly analytically computable.  But one can study various limiting solutions of these above mentioned integrals.  Let us first look into the \underline{ small $q$ with large $N$ and any arbitrary $\beta$} limiting approximation under which we get the following analytical expressions:
\begin{widetext}
\begin{eqnarray}
 \mathcal{J}_{1} (N,q,\beta):&=& \frac{1}{16}~\exp(2N^2\beta)~\Bigg\{8\Delta^{3+2N^2(q-1)\beta}~{\rm ExpIntegralE}\bigg[-\frac{1}{2}+N^2\beta(1-q),N^2\Delta^2(q-1)\bigg]\nonumber\\
 &&~~~~~~~~~~~~~~~~~~~~~~~~~~~~~~~-2^{-2N^2+(q-1)\beta}~~{\rm ExpIntegralE}\bigg[-\frac{1}{2}+N^2\beta(1-q),\frac{1}{4}N^2(q-1)\bigg] \Bigg\},\\
 \mathcal{J}_{2} (N,q,\beta):&=& 2^{\displaystyle \frac{N^2}{2}(q-1)}~\exp\left(-\frac{N^2}{4}(q-1)\right)\exp(2N^2\beta)\nonumber\\
 &&~~~~\times\Bigg\{\frac{\displaystyle \Gamma \left(\frac{1}{2} (q-1) N^2+1\right) \Gamma \left(2 (q-1) \beta  N^2+3\right)}{\displaystyle \Gamma \left(\frac{1}{2} (q-1) (4 \beta +1) N^2+4\right)}-B_{\Lambda +\frac{1}{2}}\left(2 (q-1) \beta  N^2+3,\frac{1}{2} (q-1) N^2+1\right)\Bigg\}.~~~~~~~~~
\end{eqnarray}
\end{widetext}
Next we consider the \underline{large $N$,  extremely small $\beta$} along-with the constraint  \underline{$N^2\beta$ is finite and very small} and \underline{no restriction in $q$} limiting approximation under which we get the following analytical expressions:
\begin{widetext}
\begin{eqnarray}
 \mathcal{J}_{1} (N,q,\beta):&=& \frac{1}{4 N^3 (q-1)}\Bigg[\frac{\displaystyle \sqrt{\pi } \left(\text{erfi}\left(\frac{1}{2} N \sqrt{1-q}\right)-\text{erfi}\left(\Delta  N \sqrt{1-q}\right)\right)}{\sqrt{1-q}}\nonumber\\
 &&~~~~~~~~~~~~~~~~~~~~~~~~~~~~~~~~~~+2N \Delta  \exp(-N^2\Delta ^2(q-1))-N ~\exp\left(-\frac{N^2}{4}(q-1)\right)\Bigg]\nonumber\\
 &&~~+2 \beta  N^2 \Bigg\{\frac{1}{2} \Delta ^{q+2}~ {\rm ExpIntegralE}\bigg[-\frac{q}{2},N^2 (q-1) \Delta ^2\bigg]-2^{-q-3} {\rm ExpIntegralE}\bigg[-\frac{q}{2},\frac{1}{4} N^2 (q-1)\bigg]\Bigg\},\\
 \mathcal{J}_{2} (N,q,\beta):&=& \frac{\displaystyle 2 \beta  N^2 \Gamma (q+2) \Gamma \left(\frac{1}{2} (q-1) N^2+1\right)}{\displaystyle\Gamma \left(\frac{1}{2} (q-1) N^2+q+3\right)}-2 \beta  N^2 B_{\Lambda +\frac{1}{2}}\left(q+2,\frac{1}{2} (q-1) N^2+1\right)\nonumber\\
 &&+\frac{\displaystyle (1-2 \Lambda ) \left(\frac{1}{2}-\Lambda \right)^{\frac{1}{2} N^2 (q-1)} \left(32 \Lambda  (\Lambda +2)+(2 \Lambda +1)^2 N^4 (q-1)^2+2 (2 \Lambda +1) (6 \Lambda +7) N^2 (q-1)+56\right)}{4 \left(N^2 (q-1)+2\right) \left(N^2 (q-1)+4\right) \left(N^2 (q-1)+6\right)}.~~~~~~~~~
\end{eqnarray}
\end{widetext}
These analytical results will be helpful to understand the behaviour of the sub-leading contributions which contributes to the higher genus contributions to the half-wormholes in the present context of discussion.

   Hence the full general result of SFF for the $O(N)^{q-1}$ tensor model can be expressed as:
   \begin{widetext}
\begin{eqnarray}
  {\bf SFF}
  &=&  \frac{1}{|L|N^2}\sum^{N}_{n,n',m,m'=1}\delta(\theta^{n}_{m} - \theta^{n'}_{m'}) +\underbrace{\frac{1}{|L|^{2}N^2}\sum^{N}_{n,n',m,m'=1} \delta_{n,n'}^{L}\delta_{m,m'}^{R}}_{\textcolor{red}{\rm Contribution~from ~wormhole~saddle~at~u=0}}\nonumber\\
 &&~~~~~+\underbrace{\underbrace{\frac{1}{|L|^{2}N^2}\sum^{N}_{n,n',m,m'=1} \delta_{n,n'}^{L}\delta_{m,m'}^{R}~\mathcal{J}_{1} (N,q,\beta)}_{\textcolor{red}{\rm Contribution~from ~half-wormhole~saddle~at~u\leq \frac{1}{2}}}+\underbrace{\frac{1}{|L|^{2}N^2}\sum^{N}_{n,n',m,m'=1}  \delta_{n,n'}^{L}\delta_{m,m'}^{R}~\mathcal{J}_{2} (N,q,\beta)}_{\textcolor{red}{\rm Contribution~from ~half-wormhole~saddle~at~u>\frac{1}{2}}}}_{\textcolor{blue}{\rm Contribution~from ~half-wormhole~saddle~with~|u|=1}}\nonumber\\
 &=&\frac{1}{N^2}\Bigg\{ \underbrace{ \frac{1}{|L|}\sum^{N}_{n,n',m,m'=1}\delta(\theta^{n}_{m} - \theta^{n'}_{m'})}_{\textcolor{violet}{\rm Leading~contribution}}+  \underbrace{\frac{1}{|L|^{2}}~\sum_{n,n',m,m'=1}^{N} \delta_{n,n'}^{L}\delta_{m,m'}^{R} \Bigg[1+ \mathcal{J}_{1}  (N,q,\beta)+\mathcal{J}_{2} (N,q,\beta)\Bigg]}_{\textcolor{violet}{\rm Sub-leading~contribution}}\Bigg\}\nonumber\\
 &=&\Bigg\{  \frac{1}{|L|}+  \frac{1}{|L|^{2}}+ \frac{(\mathcal{J}_{1}  (N,q,\beta)+\mathcal{J}_{2} (N,q,\beta))}{|L|^2}\Bigg\}.~~~~~~ \label{qgk}
\end{eqnarray}
\end{widetext}
Here it is important to note that:
\begin{eqnarray}
\sum_{n,n',m,m'=1}^{N} \delta_{n,n'}^{L}\delta_{m,m'}^{R}&=&\sum_{n,n'=1}^{N}\delta_{n,n'}^{L}\sum_{m,m'=1}^{N}\delta_{m,m'}^{R}\nonumber\\
&=&N_LN_R\nonumber\\
&\sim & N^2~~~({\rm where}~~N_L\sim N_R)~~~~
\end{eqnarray}
In eq \eqref{qgk} the sum over Dirac delta functions will result in four terms but since we are working in the large $N$ limit one could argue that $n = m$ due to large number of fermionic field also by summing over in the large limit the $n \not = n'$ cross terms with oscillating phases will cancel out giving rise to leading order term which is independent of $N$.  For this reason we have used the following fact:
\begin{eqnarray}
\sum^{N}_{n,n',m,m'=1}\delta(\theta^{n}_{m} - \theta^{n'}_{m'})=N^2.
\end{eqnarray}

 The first term of above equation effectively describes the late time ramp behaviour of our quantum mechanical model in the regime where non-trivial saddle points of holonomic degrees of freedom plays a crucial role. The gravitational interpretation of the sub-leading contributions appearing in \eqref{qgk} can be interpreted in terms of higher genus wormhole or equivalently in the language of topological defects as described in previous section which breaks the global symmetry and can be seen as a non-perturbative effect. We see that the behaviour of ramp/plateau of tensor model is analogous to ramp/plateau behaviour of brownian SYK where the late time behaviour of spectral form factor is of $O(1)$ governed by the non-trivial saddle point in the collective field description.
 
 Now for the \underline{ small $q$ with large $N$ and any arbitrary $\beta$} limiting approximation the SFF can further approximately reduces to the following form:
  \begin{widetext}
\begin{eqnarray}
  {\bf SFF}
 &=&\Bigg\{ \frac{1}{|L|}+ \frac{1}{|L|^{2}}\Bigg[1+ \frac{1}{16}~\exp(2N^2\beta)~\Bigg\{8\Delta^{3+2N^2(q-1)\beta}~{\rm ExpIntegralE}\bigg[-\frac{1}{2}+N^2\beta(1-q),N^2\Delta^2(q-1)\bigg]\nonumber\\
 &&~~~~~~~~~~~~~~~~~~~~~~~~~~~~~~~~~~~~~~~~~~~~~~~~~~~~-2^{-2N^2+(q-1)\beta}~~{\rm ExpIntegralE}\bigg[-\frac{1}{2}+N^2\beta(1-q),\frac{1}{4}N^2(q-1)\bigg] \Bigg\}\nonumber\\
 &&~~~~~~~+2^{\displaystyle \frac{N^2}{2}(q-1)}~\exp\left(-\frac{N^2}{4}(q-1)\right)\exp(2N^2\beta)\times\Bigg\{\frac{\displaystyle \Gamma \left(\frac{1}{2} (q-1) N^2+1\right) \Gamma \left(2 (q-1) \beta  N^2+3\right)}{\displaystyle \Gamma \left(\frac{1}{2} (q-1) (4 \beta +1) N^2+4\right)}\nonumber\\
 &&~~~~~~~~~~~~~~~~~~~~~~~~~~~~~~~~~~~~~~~~~~~~~~~~~~~~~~~~~~-B_{\Lambda +\frac{1}{2}}\left(2 (q-1) \beta  N^2+3,\frac{1}{2} (q-1) N^2+1\right)\Bigg\}\Bigg]\Bigg\}.~~~~~~ \label{qq}
\end{eqnarray}
\end{widetext}
Also  \underline{large $N$,  extremely small $\beta$} along-with the constraint  \underline{$N^2\beta$ is finite and very small} and \underline{no restriction in $q$} limiting approximation the SFF can further approximately reduces to the following form:
  \begin{widetext}
\begin{eqnarray}
  {\bf SFF}
 &=&\Bigg\{ \frac{1}{|L|}+ \frac{1}{|L|^{2}}~\Bigg[1+ \frac{1}{4 N^3 (q-1)} \Bigg\{\frac{\displaystyle \sqrt{\pi } \left(\text{erfi}\left(\frac{1}{2} N \sqrt{1-q}\right)-\text{erfi}\left(\Delta  N \sqrt{1-q}\right)\right)}{\sqrt{1-q}}\nonumber\\
 &&~~~~~~~~~~~~~~~~~~~~~~~~~~~~~~~~~~~~~~~~~~~~+2N \Delta  \exp(-N^2\Delta ^2(q-1))-N ~\exp\left(-\frac{N^2}{4}(q-1)\right)\Bigg\}\nonumber\\
 &&~~+2 \beta  N^2 \Bigg\{\frac{1}{2} \Delta ^{q+2}~ {\rm ExpIntegralE}\bigg[-\frac{q}{2},N^2 (q-1) \Delta ^2\bigg]-2^{-q-3} {\rm ExpIntegralE}\bigg[-\frac{q}{2},\frac{1}{4} N^2 (q-1)\bigg]\Bigg\}\nonumber\\
 &&~~~~~~~~~~~~~~~~~+ \frac{\displaystyle 2 \beta  N^2 \Gamma (q+2) \Gamma \left(\frac{1}{2} (q-1) N^2+1\right)}{\displaystyle\Gamma \left(\frac{1}{2} (q-1) N^2+q+3\right)}-2 \beta  N^2 B_{\Lambda +\frac{1}{2}}\left(q+2,\frac{1}{2} (q-1) N^2+1\right)\nonumber\\
 &&+\frac{\displaystyle (1-2 \Lambda ) \left(\frac{1}{2}-\Lambda \right)^{\frac{1}{2} N^2 (q-1)} \left(32 \Lambda  (\Lambda +2)+(2 \Lambda +1)^2 N^4 (q-1)^2+2 (2 \Lambda +1) (6 \Lambda +7) N^2 (q-1)+56\right)}{4 \left(N^2 (q-1)+2\right) \left(N^2 (q-1)+4\right) \left(N^2 (q-1)+6\right)}\Bigg]\Bigg\}.~~~~~~ \label{qq}
\end{eqnarray}
\end{widetext}
\begin{figure*}[htb]
	\centering
	\subfigure[\rm SFF~with~holonomy~for~$q=3$]{
		\includegraphics[width=8.5cm,height=8.7cm] {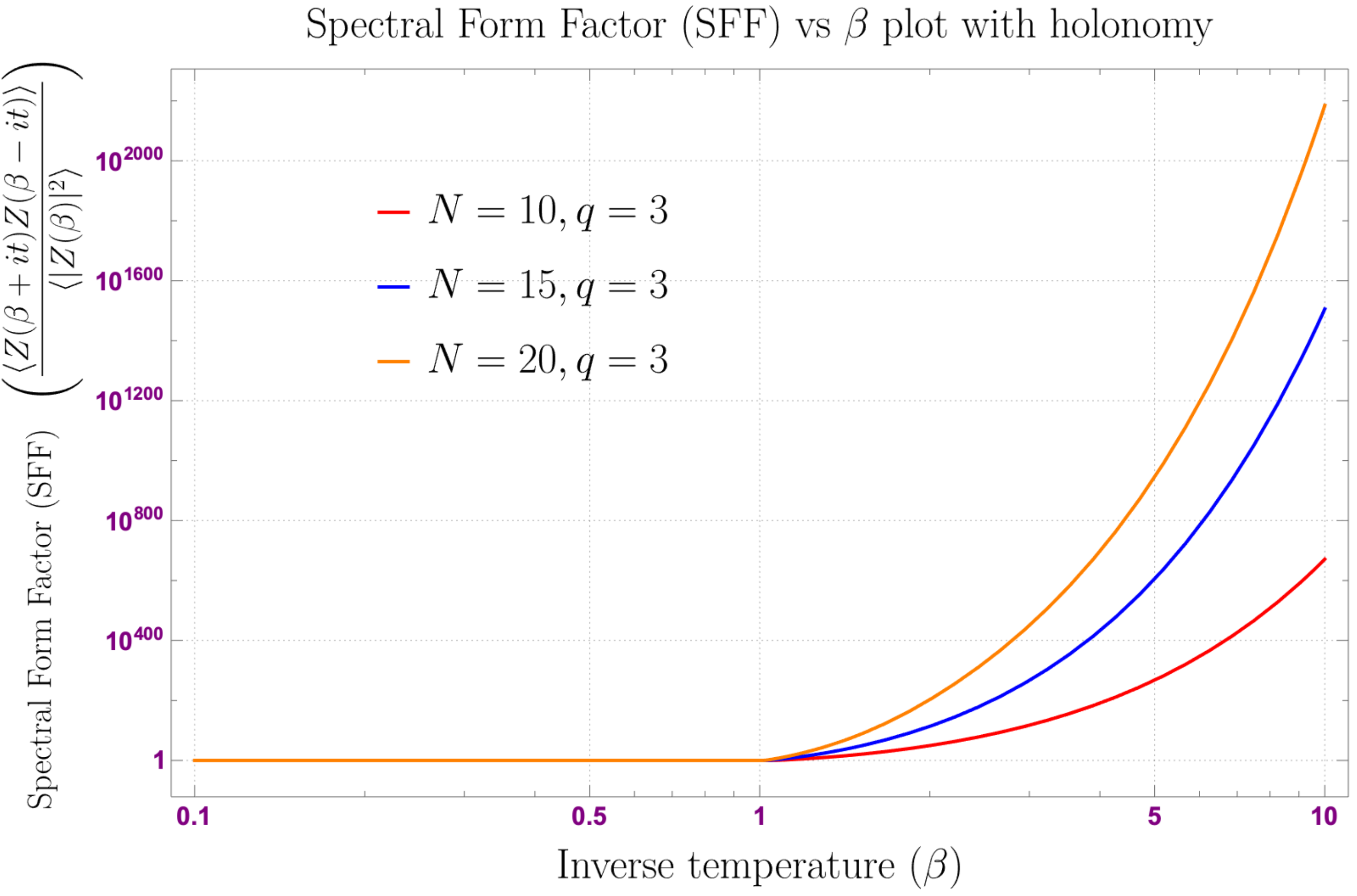}\label{SFF1a}
	}
	\subfigure[\rm SFF~with~holonomy~for~$q=4$]{
		\includegraphics[width=8.5cm,height=8.7cm] {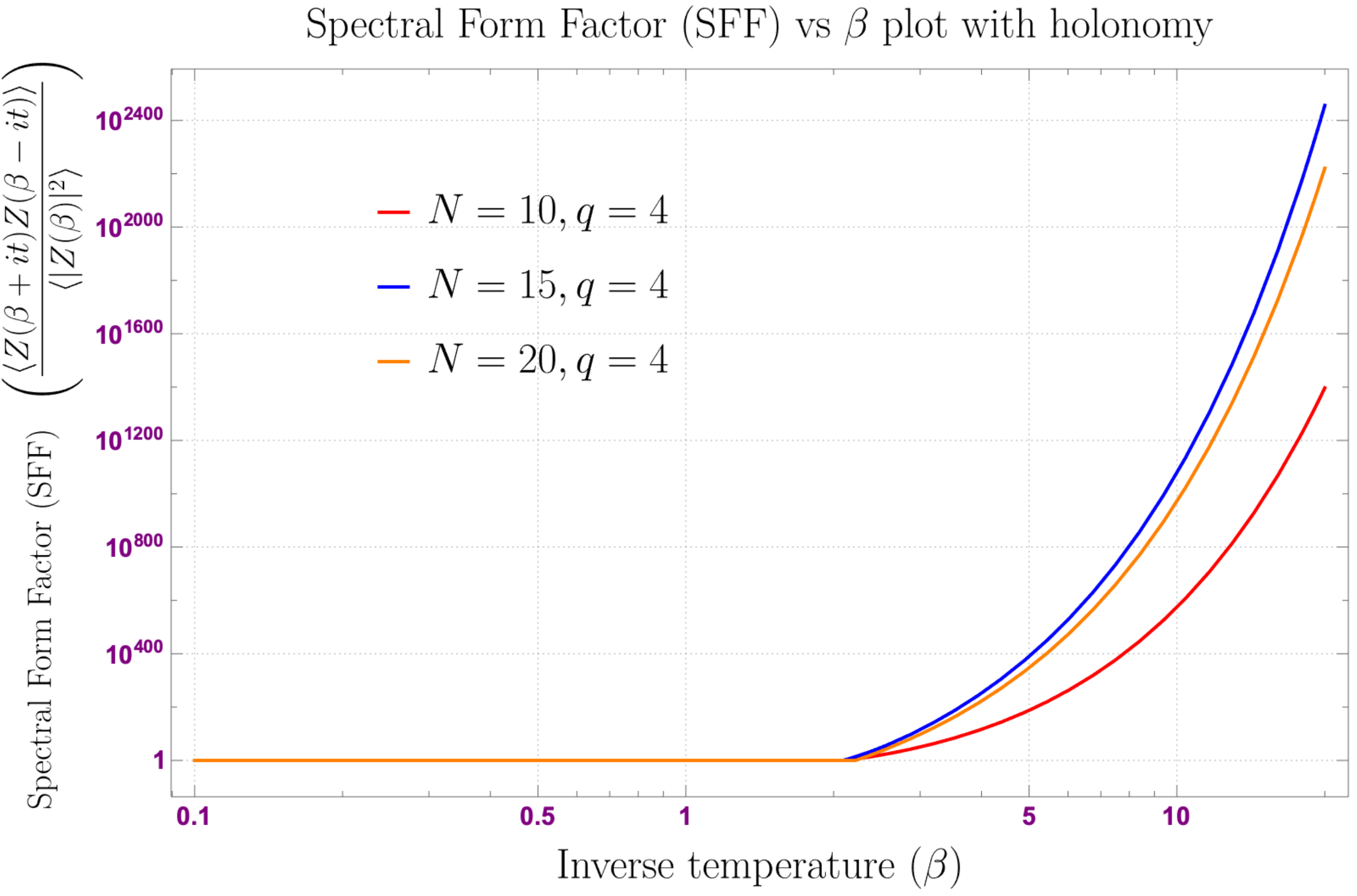}\label{SFF1b}
	}
	\subfigure[\rm SFF~with~holonomy~for~$q=5$]{
		\includegraphics[width=8.5cm,height=8.7cm] {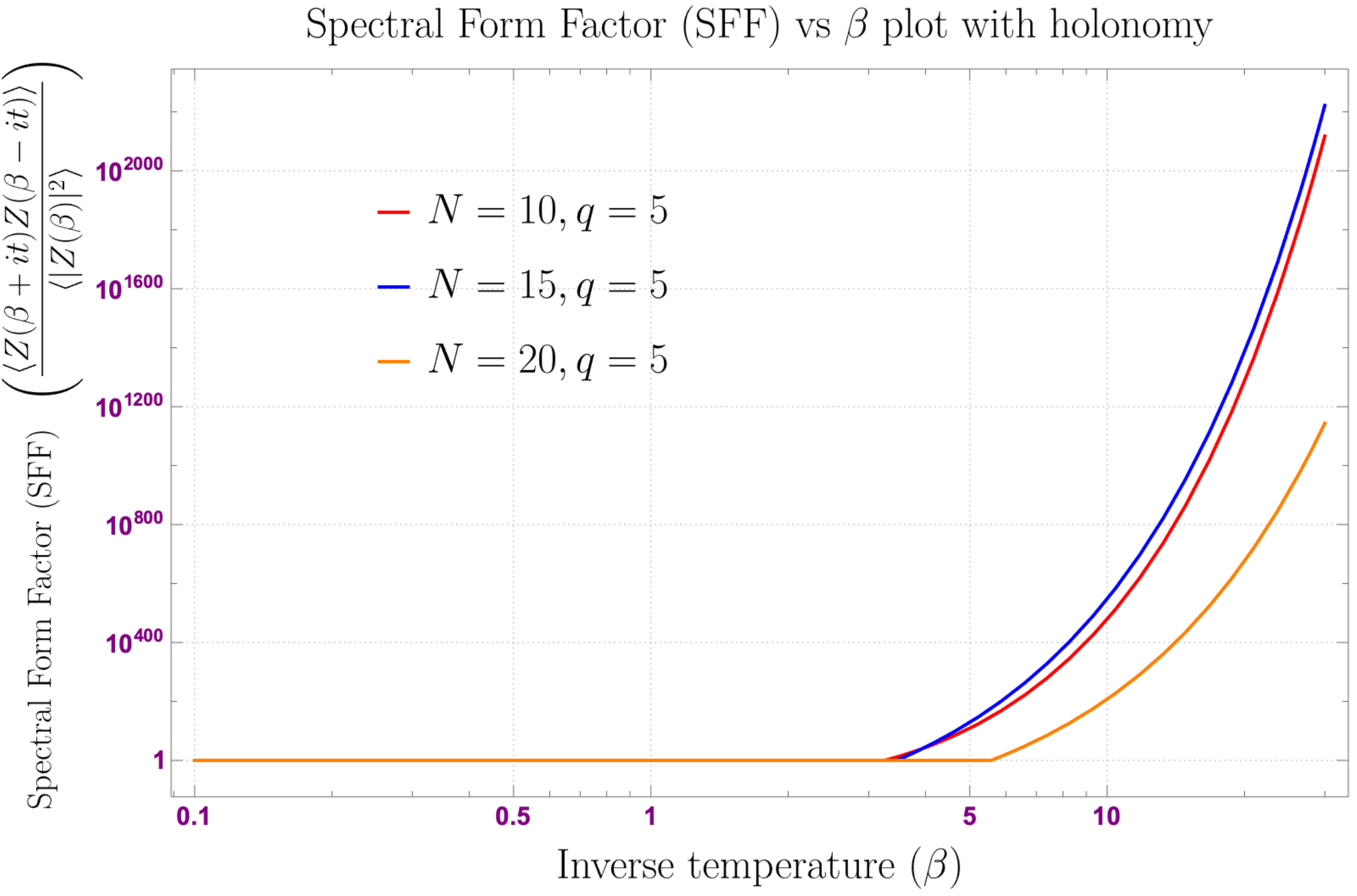}\label{SFF1c}
	}
	\subfigure[\rm SFF~with~holonomy~for~$q=6$]{
		\includegraphics[width=8.5cm,height=8.7cm] {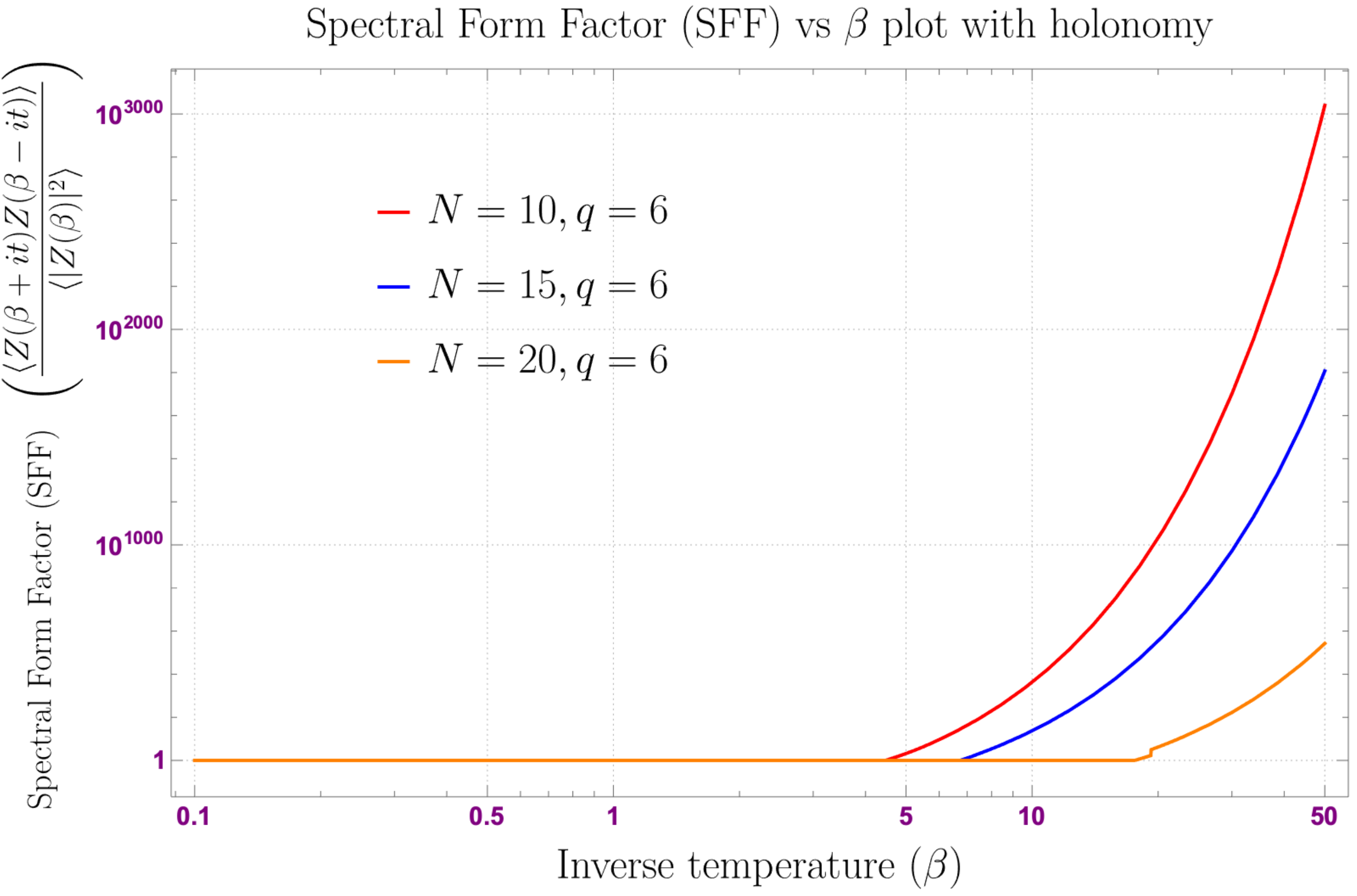}\label{SFF1d}
	}
	\caption{Behaviour of the Spectral Form Factor (SFF) with holonomy for $O(N)^{q-1}$ for tensor model with respect to inverse temperature $\beta$ for different values of the parameter $q=3$,  $q=4$, $q=5$ and $q=6$ respectively.  In these plots we have considered large $N$ parameter to be $N=10,15,20$.  }
	\label{SFF1}
\end{figure*}
\begin{figure*}[htb]
	\centering
	\subfigure[\rm SFF~with~holonomy~for~$q=4$]{
		\includegraphics[width=8.5cm,height=10.7cm] {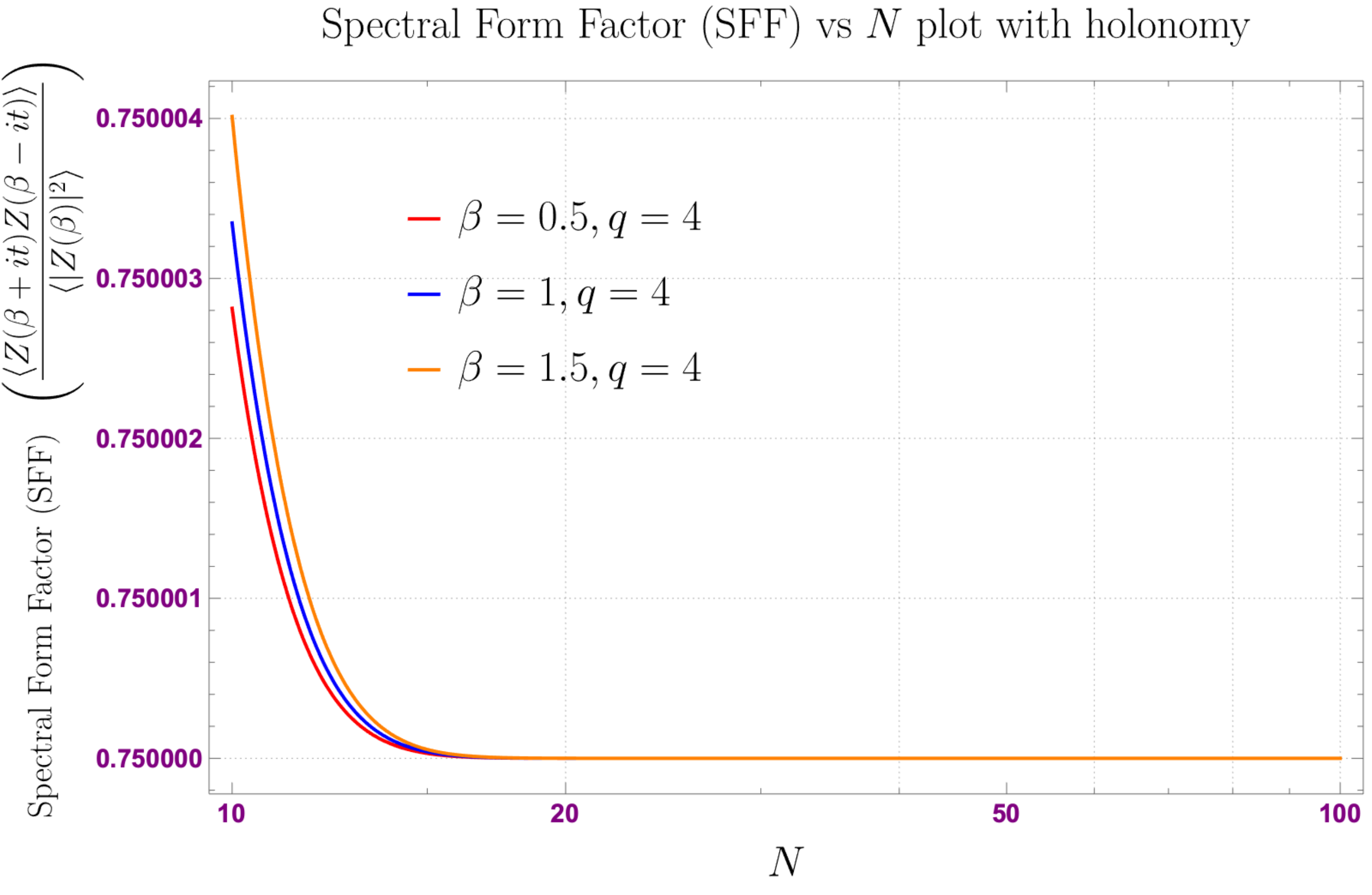}\label{SFF1e}
	}
	\subfigure[\rm SFF~with~holonomy~for~$N=10$]{
		\includegraphics[width=8.5cm,height=10.7cm] {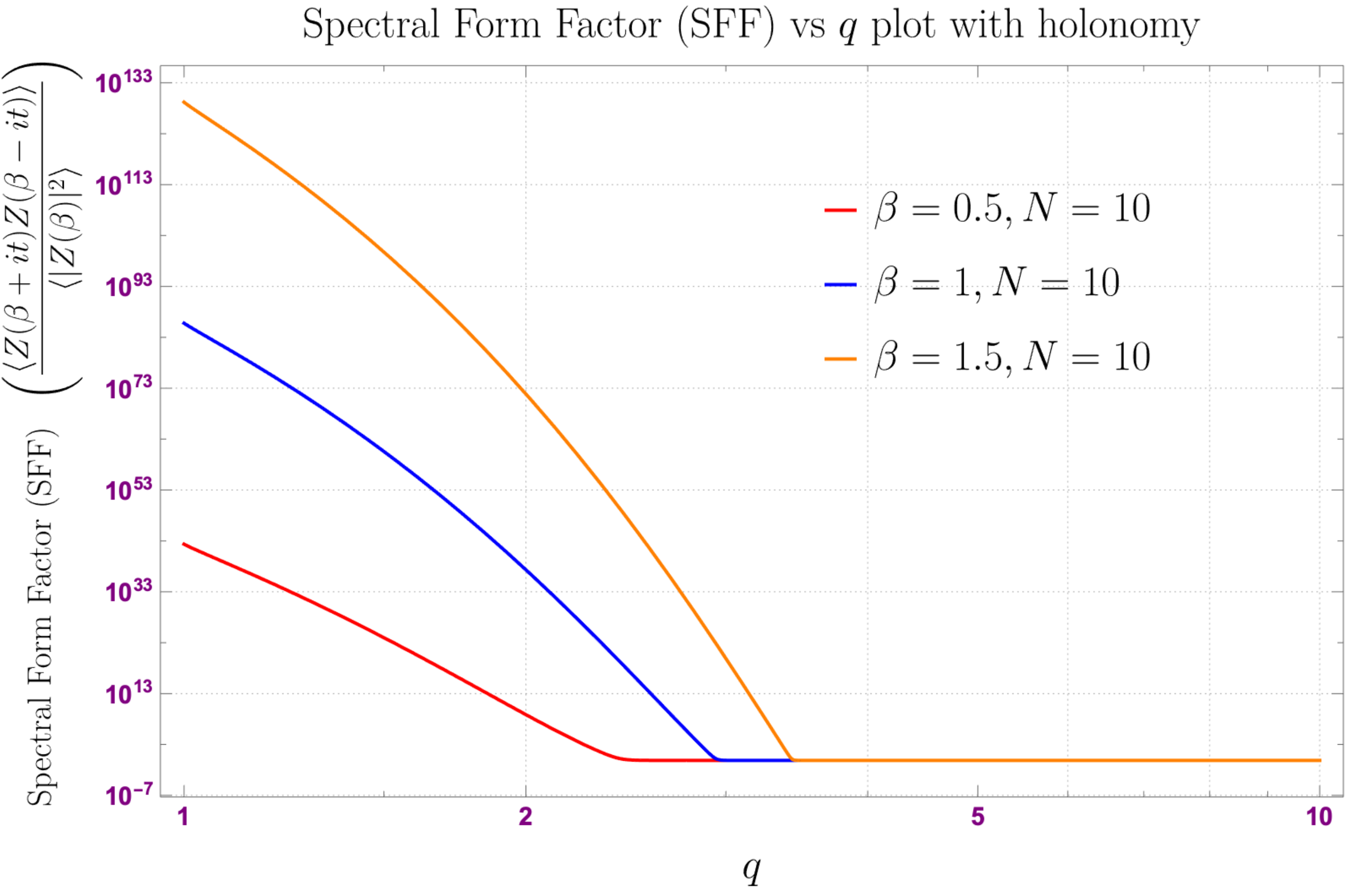}\label{SFF1f}
	}
	\caption{Behaviour of the Spectral Form Factor (SFF) with holonomy for $O(N)^{q-1}$ for tensor model with respect to $N$ and $q$.  Here for different values of the parameter $\beta=0.5,1,1.5$ respectively.   }
	\label{SFF1x}
\end{figure*}
In the regime where the dynamics of the gauge holonomy is just the identity matrix, we have seen that the dominant contribution to the partition function comes from the $N^{q-1}$ term (energy contribution) and thus reduces to the partition function of the original SYK for which we define:
\begin{widetext}
 \begin{equation}
\langle Z(\beta + it)Z(\beta - it)\rangle = \int \int d\lambda_{1}d\lambda_{2}~\langle \rho(\lambda_{1})\rangle\langle \rho(\lambda_{2})\rangle~ \exp(-\beta(\lambda_{1} + \lambda_{2}))~\exp(-i(\lambda_{1} - \lambda_{2})t)
 \end{equation}
 \end{widetext}
In this context,  the density-density correlator can be written in terms of connected and disconnected parts as follows:
\begin{widetext}
 \begin{equation}
 \langle \rho(\lambda_{1})\rho(\lambda_{2}))\rangle = \langle \rho(E)\rangle\delta(\tilde{E}) +  \langle \rho(\lambda_{1})\rangle\langle\rho(\lambda_{2})\rangle\left(1 - \frac{sin^{2}[\pi\langle\rho(E)\rangle \tilde{E}]}{[\pi\langle\rho(E)\rangle \tilde{E}]^{2}}\right)\label{mmdf}
 \end{equation}
 \end{widetext}
 where we define the density $\rho(E)$ and the newly defined variable  $\tilde{E}$ as:
 \begin{equation}
 \rho(E) \propto {\rm sinh}\left(\pi\sqrt{2} \tilde{E}\right), ~~~ \tilde{E} = \frac{c(E - E_{0})N}{\pi^{2}},~~~c=\frac{\pi^2}{q^2 j}.
 \end{equation}
 Thus the SFF gets contribution from the disconnected part and has been calculated in the ref.  \cite{Cotler:2016fpe} as:
 \begin{widetext}
 \begin{equation}
{\bf SFF}= \frac{\langle Z(\beta + it)Z(\beta - it)\rangle}{\langle |Z(\beta)|^{2}\rangle} = \frac{\beta^{3}}{|L|^2(\beta^{3} + t^{3})^{3/2}}\exp\left(-\frac{cNt^{2}}{\beta(\beta + t^{2})}\right)+g_{ramp}(t)
 \end{equation}
 \end{widetext}
when we evaluate the second term in eq \eqref{mmdf} it gives rise to ramp in the spectral form factor which grows linearly with time as follows:
\begin{widetext}
\begin{equation}\label{Sayan}
 g_{ramp}(t) =  \left\{
	\begin{array}{ll}
	 \displaystyle \frac{t}{2\pi}\exp\left[-2N s_{0} - \frac{cN}{\beta}\right] &  \frac{t}{2\pi} < e^{Ns_{0}} \\ \\
		 \displaystyle \frac{t}{2\pi}\exp\left[-2N s_{0} - \frac{cN}{\beta} - \frac{\beta}{cN}log^{2}\left(\frac{t/2\pi}{e^{Ns_{0}}}\right)\right] &  e^{Ns_{0}} < \frac{t}{2\pi}  < \frac{t_{p}}{2\pi} \\ \\
		 \displaystyle \exp\left[-Ns_{0} - \frac{3cN}{4\beta}\right] &  t_{p} < t.
	\end{array}
\right.
\end{equation}
\end{widetext}
 In the above computation in absence of the holonomy the definition of the thermofield double state will be different from the previously mentioned definition,  and it is given by the following expression:
 \begin{eqnarray}
 | \Psi \rangle_{\bf TFD} = \frac{1}{\sqrt{|L|}}\sum_{n}\exp\left(-\left(it+\frac{\beta}{2}\right)E_n\right)~|n\rangle_1 |n\rangle_2.~~~~~~~
 \end{eqnarray}
 We see that the the SFF of tensor model (without the contribution of holonomy) differs from the SFF of tensor model (with holonomy contribution) particularly in the ramp region which has $O(1)$ contributions coming from the non-trivial saddle points at late times while the later grows linearly with time also at $\beta = 0$ their are regular oscillations in the slope region when the holonomy of gauge group is identity (similar to original SYK). 
  \begin{figure}[h!]
 	\centering
 	\includegraphics[width=9cm,height=9.7cm]{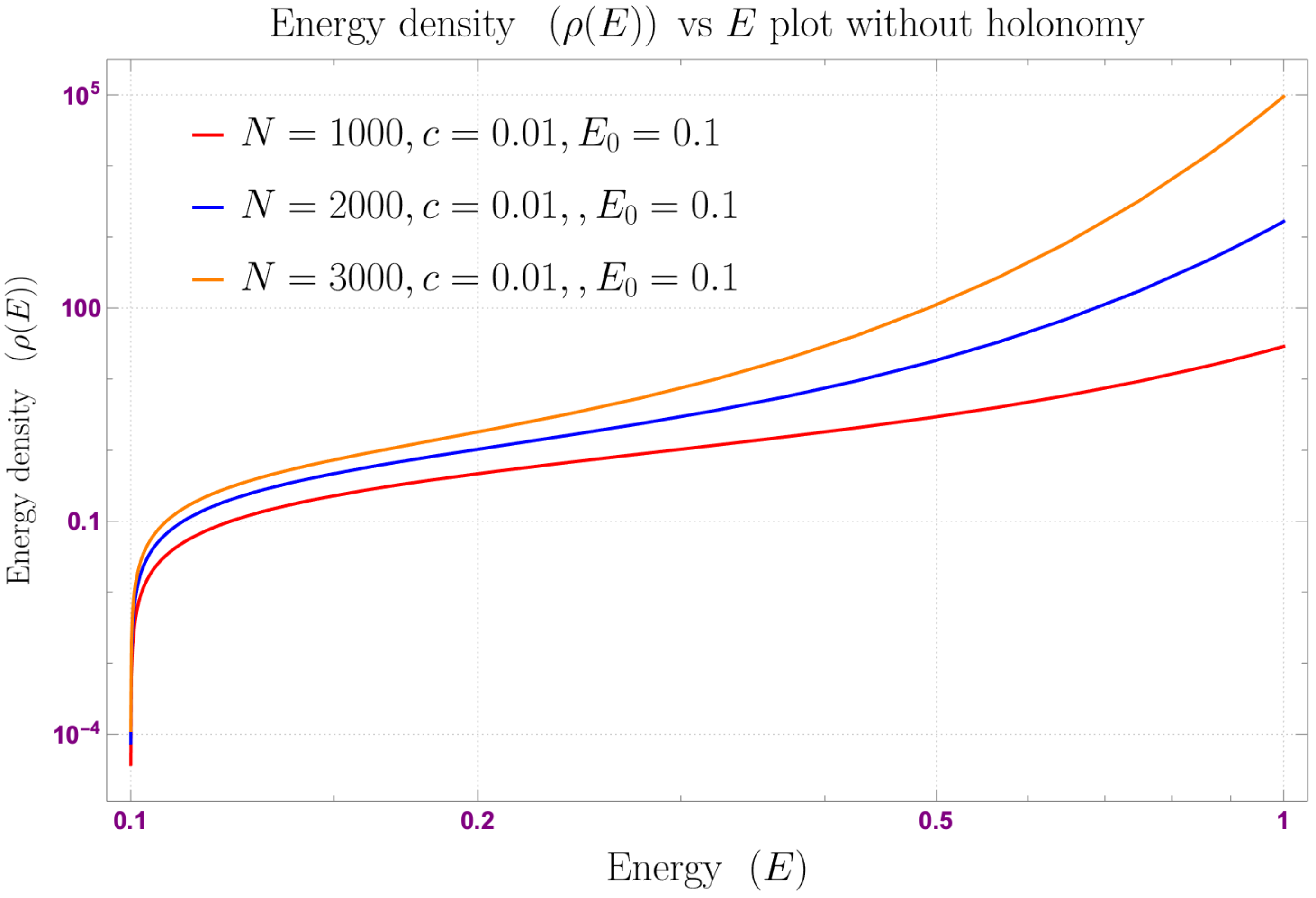}
 	\caption{Behaviour of energy density vs energy without holonomy. Here we have fixed the parameter $c=0.01$ and $E_0=0.1$ respectively.}
 \label{fig:khgg}
 \end{figure} 
  \begin{figure}[h!]
 	\centering
 	\includegraphics[width=9cm,height=9.7cm]{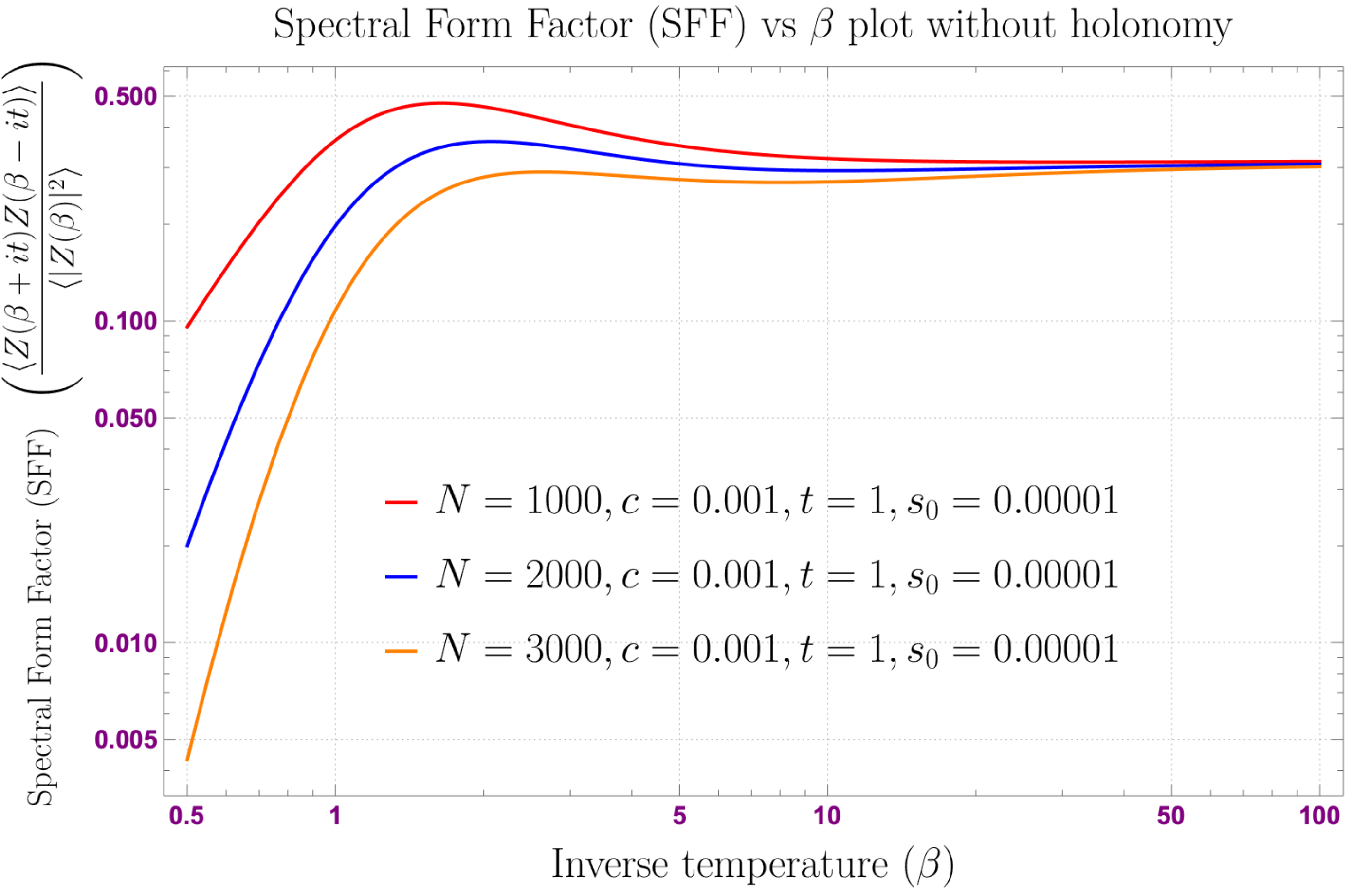}
 	\caption{Behaviour Spectral Form Factor (SFF) without holonomy vs the inverse temperature $\beta$.  Here we have fixed the time scale $t=1$, other parameters $s_0=0.00001$ and $c=0.001$ respectively.  Additionally,  we have considered the different values of the large $N$ parameter $N$ which are $N=1000,2000, 3000$ respectively.}
 \label{fig:khggc}
 \end{figure} 
\begin{figure*}[htb]
	\centering
	\subfigure[\rm SFF~with~holonomy~for~$q=4$]{
		\includegraphics[width=8.5cm,height=9.7cm] {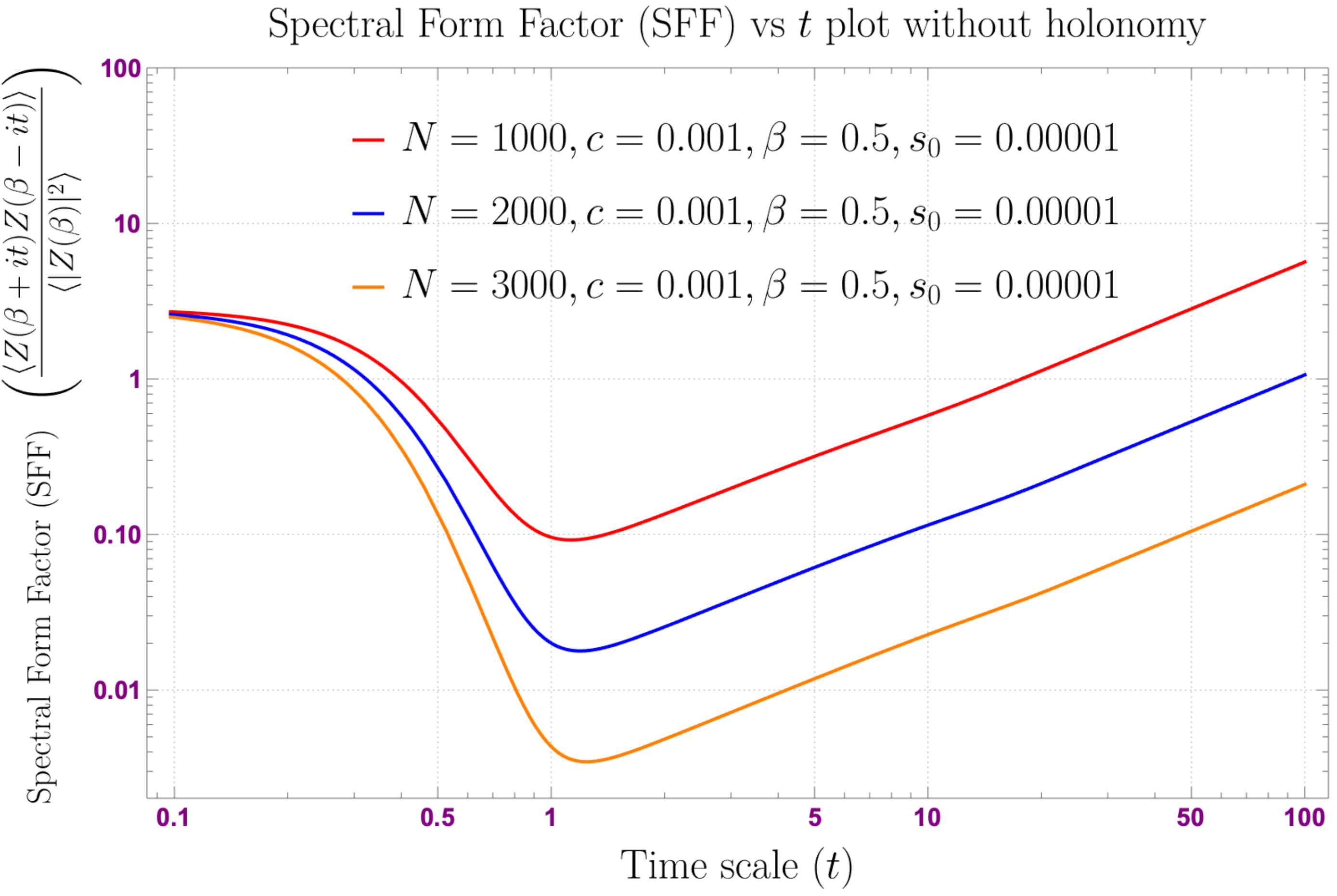}  \label{fig:ksss}
	}
	\subfigure[\rm SFF~with~holonomy~for~$N=10$]{
		\includegraphics[width=8.5cm,height=9.7cm] {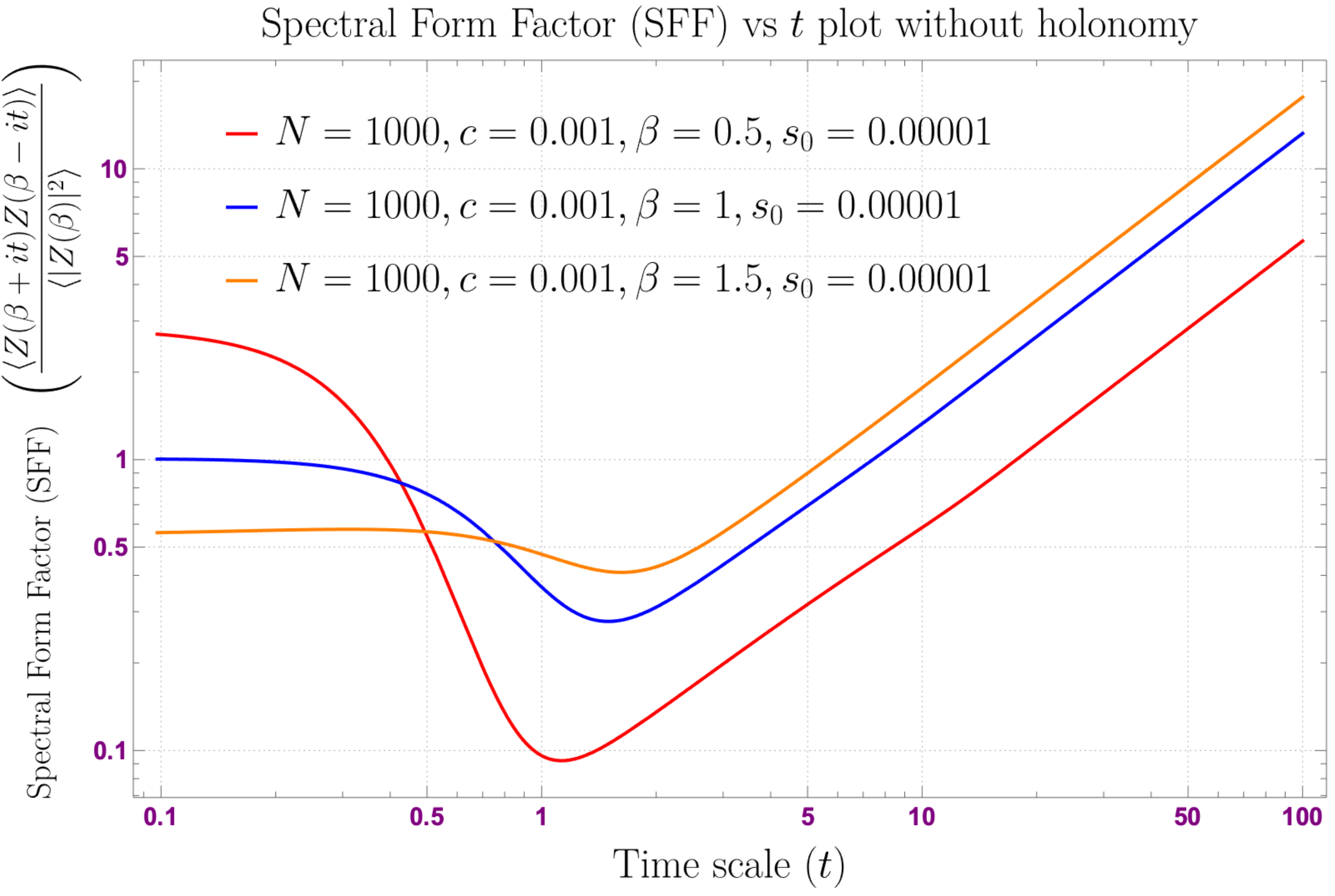} \label{fig:nnnn}
	}
	\caption{Behaviour of the Spectral Form Factor (SFF) without holonomy with respect to the time scale $t$.  Here we fix different values of the parameters $c=0.001$ and $s_0=0.00001$ respectively.  For the first plot we have fixed $\beta=0.5$ and change $N=1000,2000,3000$.  On the other hand,  for the second plot we have fixed $N=1000$ and vary $\beta=0.5, 1,1.5$ respectively.  }
	\label{SFF1xyz}
\end{figure*}

The detailed understanding of the plots with or without having holonomy in the present context are appended below point-wise:
\begin{itemize}

\item When the contributions of holonomic saddle points are taken into account in the computation we expect the theory to deviate from the usual SYK.  From \ref{SFF1a} we found such deviations for $q = 3$. from the usual SYK.  The spectral form factor for lower values of $\beta$ remains constant upto $\beta = 1$ and then have an smooth exponential rise. Here we see that the slope of the exponential rise is increasing for increasing the parameter value $N$.  The low temperature contributions specifically comes from the non-trivial saddle points of gauge holonomy which are the sub-leading contributions around $|u| = 1$ as shown in \eqref{qgk} and explicitly coming from the half-wormhole saddles.  On the other hand,  the contribution from the wormhole saddle $u=0$ and an addition contribution which is coming from the summing over Dirac Delta function over all holonomy indices of the eigen values running from $1$ to $N$.  We also see similar kind of behaviour for different values of $N$ except from the slope that we have mentioned.  However,  SFF for larger values of $N$ has more rapid growth than compared to smaller values of $N$.  In \ref{SFF1b},  we have shown a slight variance where the spectrum for $N = 15$ which has slightly more rapid growth as compared to $N = 20$.  As we go towards larger and larger values of $q$ we observer that SFF with lowest values $N$ takes over and has much more rapid growth as compared to larger values of $N$,  as explicitly shown in fig \ref{SFF1c} and \ref{SFF1d} respectively.  Also,  when we increase the value of $q$ the leading order contribution dominates for larger values of $\beta$ before it starts rising for $N = 10$ as compared to $N = 20$ and $15$. However the mirror image of the SFF with respect to inverse temperature $\beta$ has similar behaviour with that of SYK where at low temperatures the SFF has a slope region and then intersects with the plateau expect that it has no ramp which grows linearly with time.  We observe similar kind of feature in the Brownian SYK where the plateau has $O(1)$ contributions coming from the non-trivial saddle points.  In the $O(N)^{q - 1}$ tensor model the non-trivial half-wormhole saddle point contributions are dominant at low temperatures (large $\beta$) and constant plateau type of contribution is dominant at high temperature (low $\beta$).  Apart from the constant part and the rising part of the plots for SFF,  the significant deformation which is describing by the dominant behaviour of worm hole saddle contribution over the half-wormhole saddles for increasing values of the parameter $q$ along-with the increasing value of the large $N$ parameter is one of the highlighting findings from these plots for $O(N)^{q-1}$ tensor models with holonomy. 
 
\item In fig \ref{SFF1e},  we have plotted the behaviour of SFF with respect the large $N$ parameter for a fixed value of $q$,  which we we done at $q=4$ and for different values of inverse temperature $\beta$ which covers low, intermediate and high temperature behaviour in a single plot.   From this plot we observe that for all the values of $\beta$ the SFF has similar kind of behaviour, where for smaller values of $N$ SFF has a decaying slop region coming from the non-trivial holonomic saddle points coming from half-wormhole contributions and then quickly saturates between $ 10 < N < 20$, which is coming from the wormhole saddle point contribution along with the summing over Dirac Delta function contribution over all holonomy indices.  We additionally observe that the SFF for $\beta = 0.5$ is little more steeper than the contribution coming from $\beta = 1$.  A similar kind of behaviour is observed between $\beta = 1$ and $1.5$ also.

\item In fig \ref{SFF1f},  we have discussed the behaviour of spectral form factor with respect to the parameter $q$ for fixed value of $N = 10$ and three different values of $\beta$,  which basically cover the low,  intermediate and high temperature behaviour respectively.  We observe that for $\beta = 0.5$ the SFF decreases linearly and then saturates around $q = 3$.  We observe similar kind of behaviour for $\beta = 1$ but now has larger value of SFF for small values of $q$ which then reach a saturation point around $q=4$. Thus we observe that as we increase $\beta$ the SFF has large values for small $q$ compared to smaller values of $\beta$.

\item In fig \ref{fig:khgg},  we have observed the behaviour of energy density vs energy in the absence of gauge holonomy for fixed values of parameters $c = 0.01$ (coupling) and $E_{0} = 1.1$ and for three different values of $N$. We see that energy density shows a sudden jump in its value for $E = 0.1$ and from their on increases quit linearly with respect to $E$ for some values of $E$ and then has an exponential rise.  We observe that for larger values of $N$ the exponential rise of energy density starts on early i.e for smaller values of $E$ as compared to lower values of $N$. 

\item We see that in the large $N$ limit the ramp described above saturates and intersects the plateau which has a height of ${\cal O}(L)$.  The three different graphs with different values $N$ seems to merge for large values of inverse $\beta$ as shown in fig \ref{fig:khggc}. The non-zero value of plateau is an indication discreteness of energy spectrum and is plotted by taking an disordered average over the coupling.  For fixed choice of coupling the slope we expect the late time behaviour to have massive fluctuations which is not shown in here.

\item In the absence of holonomy we know that the large $N$ behaviour of tensor model is indistinguishable to that of SYK.  In figure \ref{fig:nnnn},  we study the behaviour of spectral form factor particularly in the absence of holonomy for three different values of $\beta$.  We see that at early times the Fig has a slope which decreases rapidly and is described by the simple Schwarzian action after some time the intersects with ramp which increases linearly with time. The region where transition from slope to ramp takes place is effectively described by spin-glass phase that is present in the Sachdev-Ye model.  As we decrease the value $\beta$ one finds that their is an increase in the slope portion which decreases more rapidly at earlier values of time. In Fig \ref{fig:ksss},  we see similar behaviour for three different values of $N$ where region of slope increases for larger values of $N$ and more rapid fall as compared to smaller values of $N$. 
\end{itemize}  
 \section{\textcolor{Sepia}{\textbf{ \Large Wormhole and half-wormhole energetics}}}\label{WE}
 In this section,  our prime objective is to find the contributions from the wormhole and half-wormholes in particularly in the energetics,. This is described by mainly by averaged energy,  averaged free energy and averaged entropy.  In this context,  to take the average over all of these thermodynamic quantities which describes the energetics,  we have to take the average over the holonomy dependent thermofield double state for $O(N)^{q-1}$ tensor model,  which we have explicitly defined in the equation \eqref{HTFD}. This is exactly analogous approach that we follow in the context of Statistical Mechanics to compute the statistical ensemble averages of the above mentioned thermodynamic quantities.  
 
 To study the wormhole and half-wormhole energetics we will follow the following strategies,  which we have stated point-wise in the following:
 \begin{enumerate}
 \item First of all,  we take the derived expression for the energy,  free energy and entropy from the holonomy dependent partition functions (see equation \eqref{Sayang}) for $O(N)^{q-1}$ tensor model as stated in equation \eqref{Sayan2},  \eqref{Sayan3} and \eqref{Sayan4}.
 
 \item Then using the definition of the holonomy dependent thermofield double state for $O(N)^{q-1}$ tensor model as stated in equation \eqref{HTFD} we derive the averaged contributions from the energetics by integrating over the holonomies around the wormhole saddle $u=0$ and from half-wormhole saddle point contribution appearing around $|u|=1$.
 
 \item   Here the integrating over holonomies can be performed by taking the integration over the Haar measure explicitly in a specified range of $u$ which can cover both the information coming from the saddles around $u=0$ and $|u|=1$.
 \end{enumerate}
 
 After following the above mentioned steps we get the following contributions from the averaged energetics:
 \begin{widetext}
\begin{equation}\label{Sayan2a}
 \langle E(\beta)\rangle =\large \left\{
	\begin{array}{ll}
	 \displaystyle 0~~~~~~~~~~~~~~~ & \mbox{if } u= 0 \\ 
		 \displaystyle N^2\beta~\int^{\frac{1}{2}}_{u=\Delta>0}du~u^{q+1}=\frac{N^2\beta}{q+2}\bigg[\frac{1}{2^{q+2}}-\Delta^{q+2}\bigg] & \mbox{if } u \leq \frac{1}{2} \\ 
		 \displaystyle N^2\beta~\int^{1}_{u=\frac{1}{2}+\Lambda(>0)}du~u^{q+1}=\frac{N^2\beta}{q+2}\bigg[1-\left(\frac{1}{2}+\Lambda\right)^{q+2}\bigg], & \mbox{if } u > \frac{1}{2}.
	\end{array}
\right.
\end{equation}
\begin{equation}\label{Sayan3a}
 \langle F(\beta)\rangle =\large \left\{
	\begin{array}{ll}
	 \displaystyle 0~~~~~~~~~~~~~~~ & \mbox{if } u= 0 \\ 
		 \displaystyle \frac{N^2}{2}~\int^{\frac{1}{2}}_{u=\Delta>0}du~\left[\frac{1}{\beta}(q-1)u^4-2~u^{q+1}\right]\\
		 \displaystyle= \frac{N^2}{2}~\left[\frac{1}{5\beta}(q-1)\left\{\frac{1}{32}-\Delta^5\right\}-\frac{2}{q+2}\left\{\frac{1}{2^{q+2}}-\Delta^{q+2}\right\}\right] & \mbox{if } u \leq \frac{1}{2} \\  \\
		 \displaystyle N^2~\int^{1}_{u=\frac{1}{2}+\Lambda(>0)}du~\Bigg\{u^{q+1}+\frac{u^2}{4\beta}(q-1)\bigg[\frac{1}{2}-\ln 2-\ln(1-u)\bigg]\Bigg\}\\
	 \displaystyle =N^2~\Bigg[\frac{1}{q+2}\bigg\{1-\left(\frac{1}{2}+\Lambda\right)^{q+2}\bigg\} +\frac{(q-1)}{576\beta} \bigg\{-48 \ln 2\\
	 \displaystyle-5 (2 \Lambda -1) (4 \Lambda  (\Lambda +2)+7)+6 (2 \Lambda +1)^3 \log (2 \Lambda +1)\bigg\}\Bigg] 	  & \mbox{if } u > \frac{1}{2}.
	\end{array}
\right.
\end{equation}
\begin{equation}\label{Sayan4a}
\langle S(\beta)\rangle =\large \left\{
	\begin{array}{ll}
	 \displaystyle 0 & \mbox{if } u= 0 \\ \\ \\
		 \displaystyle \frac{N^2}{2}~\int^{\frac{1}{2}}_{u=\Delta>0}du~\left[(1-q)u^4+2\beta~u^{q+1}\right]\\
		 \displaystyle -N^2\beta~\ln\frac{\beta}{N^{q-3}p}~\int^{\frac{1}{2}}_{u=\Delta>0}du~u^{q+1}\\
		 \displaystyle= \frac{N^2}{2}~\bigg[\frac{(1-q)}{5}\bigg\{\frac{1}{32}-\Delta^5\bigg\}+\frac{2\beta}{q+2}\left\{\frac{1}{2^{q+2}}-\Delta^{q+2}\right\}\bigg]  & \mbox{if } u \leq \frac{1}{2} \\ \\ \\
		 \displaystyle N^2~\int^{1}_{u=\frac{1}{2}+\Lambda(>0)}du~\Bigg\{-\beta u^{q+1}+\frac{u^2}{4}(1-q)\bigg[\frac{1}{2}-\ln 2-\ln(1-u)\bigg]\Bigg\}\\
		 \displaystyle ~~~~~~~~~~~~~~~~~~~~~~~~~ -N^2\beta~\ln\frac{\beta}{N^{q-3}p}~\int^{1}_{u=\frac{1}{2}+\Lambda(>0)}du~u^{q+1}\\
		 \displaystyle= N^2~\Bigg\{-\frac{\beta}{q+2}\left(1+\ln\frac{\beta}{N^{q-3}p}\right)\bigg[1-\left(\frac{1}{2}+\Lambda\right)^{q+2}\bigg]+\frac{(q-1)}{576} \bigg[-48 \ln 2\\
	 \displaystyle-5 (2 \Lambda -1) (4 \Lambda  (\Lambda +2)+7)+6 (2 \Lambda +1)^3 \log (2 \Lambda +1)\bigg] \Bigg\} & \mbox{if } u > \frac{1}{2}.
	\end{array} 
\right.
\end{equation}
\end{widetext}
whereat last we compute the contribution from the averaged entropy function at the large $N$ to understand how exactly in the wormhole saddle at $u=0$ and in the half-wormhole saddle around $|u|=1$ contribute in the present context.

After doing these computation we have to sum over all the contributions coming from the saddles at $u=0$ and around $|u|=1$ which finally give rise to following full contribution in the ensemble averaged energetics:
 \begin{widetext}
\begin{eqnarray}\label{Sayan2a}
 \langle E(\beta)\rangle_{\rm full} =\langle E(\beta)\rangle_{\textcolor{red}{u=0}}+\langle E(\beta)\rangle_{\textcolor{blue}{u\leq \frac{1}{2}}}+\langle E(\beta)\rangle_{\textcolor{violet}{u>\frac{1}{2}}}&=&\frac{N^2\beta}{q+2}\bigg[1+\frac{1}{2^{q+2}}-\Delta^{q+2}-\left(\frac{1}{2}+\Lambda\right)^{q+2}\bigg], 
\\ \label{Sayan3a}
  \langle F(\beta)\rangle_{\rm full} =\langle F(\beta)\rangle_{\textcolor{red}{u=0}}+\langle F(\beta)\rangle_{\textcolor{blue}{u\leq \frac{1}{2}}}+\langle F(\beta)\rangle_{\textcolor{violet}{u>\frac{1}{2}}}&=&\frac{N^2}{2}~\left[\frac{1}{5\beta}(q-1)\left\{\frac{1}{32}-\Delta^5\right\}-\frac{2}{q+2}\left\{\frac{1}{2^{q+2}}-\Delta^{q+2}\right\}\right]\nonumber\\
	&& \displaystyle+N^2~\Bigg[\frac{1}{q+2}\bigg\{1-\left(\frac{1}{2}+\Lambda\right)^{q+2}\bigg\} +\frac{(q-1)}{576\beta} \bigg\{-48 \ln 2 \nonumber\\
	&& \displaystyle-5 (2 \Lambda -1) (4 \Lambda  (\Lambda +2)+7)+6 (2 \Lambda +1)^3 \log (2 \Lambda +1)\bigg\}\Bigg],\\
 \langle S(\beta)\rangle_{\rm full} =\langle S(\beta)\rangle_{\textcolor{red}{u=0}}+\langle S(\beta)\rangle_{\textcolor{blue}{u\leq \frac{1}{2}}}+\langle S(\beta)\rangle_{\textcolor{violet}{u>\frac{1}{2}}}&=& \frac{N^2}{2}~\bigg[\frac{(1-q)}{5}\bigg\{\frac{1}{32}-\Delta^5\bigg\}+\frac{2\beta}{q+2}\left\{\frac{1}{2^{q+2}}-\Delta^{q+2}\right\}\bigg]  \nonumber\\
&& +N^2~\Bigg\{-\frac{\beta}{q+2}\left(1+\ln\frac{\beta}{N^{q-3}p}\right)\bigg[1-\left(\frac{1}{2}+\Lambda\right)^{q+2}\bigg]+\frac{(q-1)}{576} \bigg[-48 \ln 2\nonumber\\
	&& \displaystyle-5 (2 \Lambda -1) (4 \Lambda  (\Lambda +2)+7)+6 (2 \Lambda +1)^3 \log (2 \Lambda +1)\bigg] \Bigg\}
 \end{eqnarray}
\end{widetext}
 \begin{figure}[htb]
 	\centering
 	\includegraphics[width=13cm,height=8.7cm]{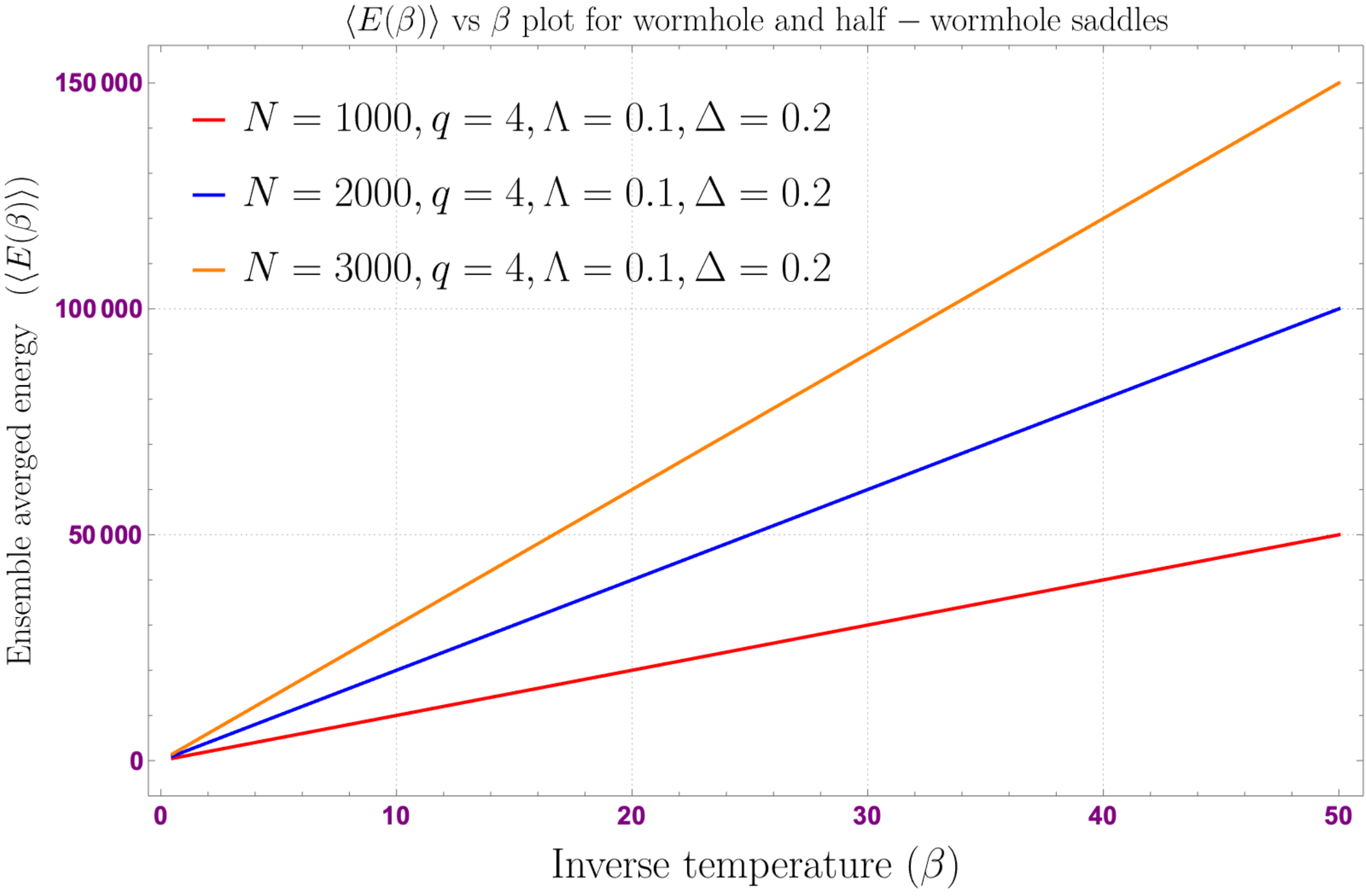}
 	\caption{High and low temperature behaviour of the ensemble averaged energy $\langle E(\beta)\rangle$ at large $N$ with respect to the inverse temperature $\beta$.  Here we have fixed three parameters $q=4$, $\Delta=0.1$ and $\Lambda=0.2$.  We have shown the behaviour for three different large $N$ values which we have taken $N=1000$, $N=2000$ and $N=3000$.}
 \label{fig:E1}
 \end{figure} 
 
  \begin{figure}[htb]
 	\centering
 	\includegraphics[width=9cm,height=8.7cm]{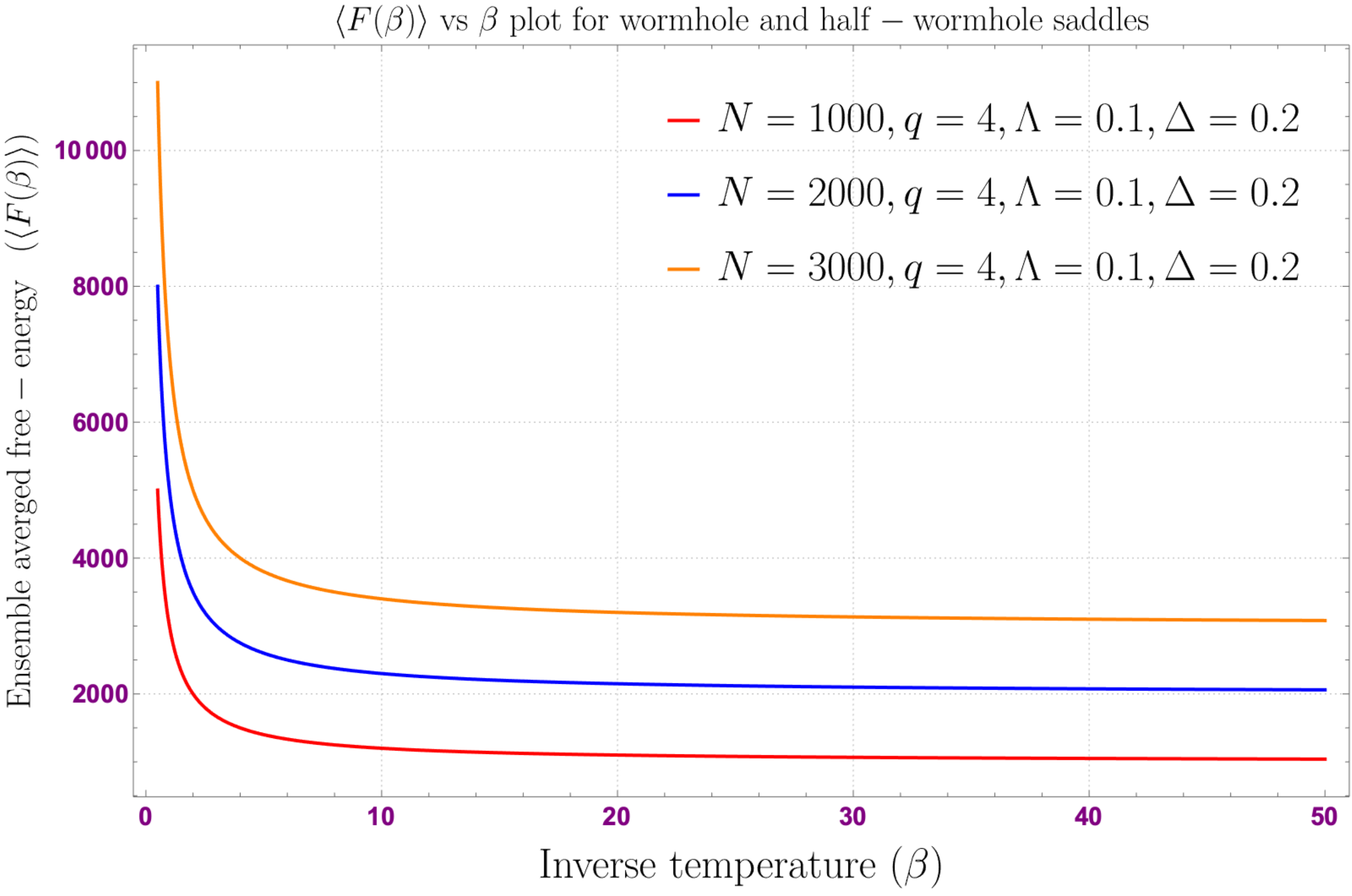}
 	\caption{High and low temperature behaviour of the ensemble averaged free-energy $\langle F(\beta)\rangle$ at large $N$ with respect to the inverse temperature $\beta$.  Here we have fixed three parameters $q=4$, $\Delta=0.1$ and $\Lambda=0.2$.  We have shown the behaviour for three different large $N$ values which we have taken $N=1000$, $N=2000$ and $N=3000$.}
 \label{fig:E2}
 \end{figure} 
 
  \begin{figure}[htb]
 	\centering
 	\includegraphics[width=9cm,height=8.7cm]{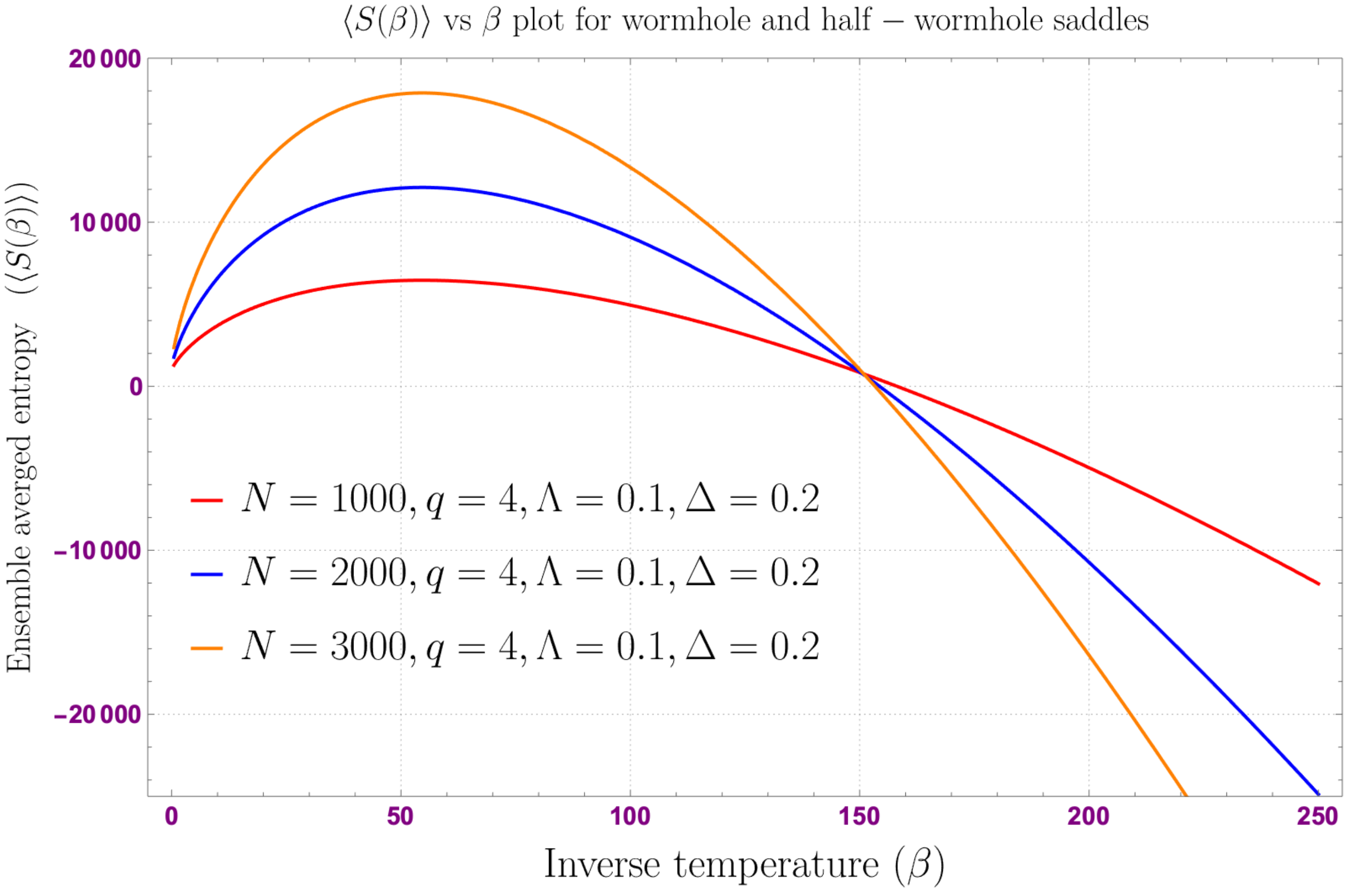}
 	\caption{High and low temperature behaviour of the ensemble averaged entropy $\langle S(\beta)\rangle$ at large $N$ with respect to the inverse temperature $\beta$.  Here we have fixed three parameters $q=4$, $\Delta=0.1$ and $\Lambda=0.2$.  We have shown the behaviour for three different large $N$ values which we have taken $N=1000$, $N=2000$ and $N=3000$.}
 \label{fig:E3}
 \end{figure} 
 From the above mentioned computations we have found the following crucial facts:
 \begin{itemize}
 \item We can explicitly see that the saddle coming from wormhole at $u=0$ not directly contributing to the energetics.
 
 \item But most importantly,  the half-wormhole saddles around $|u|=1$ which is actually coming from $u\leq \frac{1}{2}$ and $u>\frac{1}{2}$ non-trivially contribute to the ensemble averaged energetics in the present computation from the $O(N)^{q-1}$ tensor model.
 
 \item We have found that the ensemble averaged energy $\langle E(\beta)\rangle$ at large $N$ linearly grows with $\beta=1/T$,  which further implies that at very high temperature regime this particular contribution fall linearly with the temperature and in the low temperature regime it will grow linearly with temperature.  The both high and low temperature behaviour of the ensemble averaged energy $\langle E(\beta)\rangle$ at large $N$ with respect to the inverse temperature $\beta$ is explicitly shown in the figure~\ref{fig:E1},  which is quite consistent with the expectation from our computed result after summing over all contributions from saddles.

\item We also have found that the ensemble averaged free energy $\langle F(\beta)\rangle$ at large $N$ varies as $\displaystyle \left(C+\frac{D}{\beta}\right)$,  where $C$ and $D$ are the constant factors which will depend on the parameters $N$, $q$, $\Delta$ and $\Lambda$.  From the obtained feature it is expected that very low temperature regime the first term $C$ dominates over the second term $\displaystyle\frac{D}{\beta}$ and show a constant behaviour with respect to temperature.  On the other hand,  in the high temperature regime the second term $\displaystyle\frac{D}{\beta}$ dominates over the first constant term $C$ and show a linearly increasing behaviour with temperature.  The both high and low temperature behaviour of the ensemble averaged free energy $\langle F(\beta)\rangle$ at large $N$ with respect to the inverse temperature $\beta$ is explicitly shown in the figure~\ref{fig:E2},  which shows the competition between the two obtained contributions and we found that the overall behaviour is quite consistent with the expectation from our computed result after summing over all contributions from saddles. 

\item Last but not the least,  we have found that the ensemble averaged entropy $\langle S(\beta)\rangle$ at large $N$ varies as $\displaystyle \left(A+\beta\left[B-M\ln\beta\right]\right)$,  where $A$, $B$ and $M$ are the constant factors which will depend on the parameters $N$, $q$, $\Delta$ and $\Lambda$.  From the obtained feature it is expected that very low temperature regime the last term $\beta\log \beta$ and second term $\beta$ dominates over the first term $A$ and show a rising behaviour with respect to temperature.  On the other hand,  in the high temperature regime the first constant term $A$ and the last term $\beta\log \beta$ dominates over the second term $\beta$ and show a constant and then a falling behaviour with temperature.   The both high and low temperature behaviour of the ensemble averaged entropy $\langle S(\beta)\rangle$ at large $N$ with respect to the inverse temperature $\beta$ is explicitly shown in the figure~\ref{fig:E3},  which shows the competition between the three obtained contributions and we found that the overall behaviour is quite consistent with the expectation from our computed result after summing over all contributions from saddles.  We all known growing behaviour actually the measure of the disorder which visible in $\beta<50$ regime from the plot.  But after crossing this region it reaches a maximum value and the from the plot we found that the disorder reduces and and the system we are studying goes to more and more towards the ordered phase.  This depicted by the falling behaviour within $60<\beta<150$ including going towards the negative value for $\beta>150$.  This study shows that at very low temperature the system is the disordered phase, in the intermediate temperature reaches a maximum value and for very low temperature the system started moving towards the ordered phase in the present context of discussion.  Additionally it is important to note that,  in this figure we also have found a crossover at $\beta=150$ at some positive averaged value of the entropy function.  Before the crossover in $\beta>150$ we have found that as we increase the values of the large $N$ parameter considering $N=1000$,  $N=2000$ and $N=3000$ slope of the curve decrease with increasing the value of $N$.  After the crossover in the region $\beta>150$ where we have the negative entropy contribution we have found the as we increase the value of the large $N$ parameter fall in the entropy function is more and more faster with inverse temperature $\beta$.
\item We see that $u=0$ saddle doesn't contribute in the ensemble averaged entropy where as the half -wormhole saddles around $u=1$ especially in the region $u \leq \frac{1}{2}$ and $u > \frac{1}{2}$ contributes non-trivially to ensemble energetics which could be seen in figure \ref{fig:E3} where in the large $N$ limit the ensembled average entropy first rises around $u \leq \frac{1}{2}$ and then falls around $u > \frac{1}{2}$ due to non-perturbative effects in the gravitational path integral.
\item Quantum mechanically the figure \ref{fig:E3} corresponds to the non-perturbative fluctuations that originates from the discreteness of the energy spectrum. In the bulk these are hints of the higher genus topology that are not compatible with the global symmetry of the gravitational system. In sec \ref{TD} we have pointed the similar scenario where at short distances the wormhole must be equipped with topological defects that give rise to new cobordism classes that were not previously connected and are responsible for breaking of global symmetry, we showed that these objects are exactly the half-wormholes coming from the non-trivial saddle points of the holonomy whose contribution factorizes the partition function of the tensor model. In De-sitter space this is analogous to $S_{2} \times S_{2}$ Nariai geometry whose non-perturbative contribution in the gravitational path integral give rise to a massive fluctuation due to which de Sitter space turns itself "inside out".
 \end{itemize}
\section{\textcolor{Sepia}{\textbf{ \Large SYK vs. Tensor model in the framework of without averaging}}}\label{VV}
Though in the large $n$ limit the correlation function and thermodynamics of the tensor model with $O(N)^{q-1}$ gauge symmetry are same as that of the SYK but they differ in their bulk description in various ways which are as follows,
\begin{itemize}
\item The broken conformal diffeomorphism of $G^{SYK}(t)$ produces zero modes in the extreme low energy limit. However \eqref{bm} has new light model governed by effective sigma model and admits a bulk interpretation of gauge field propogating in $ADS_{2}$ background.
\item When taken average over the coupling SYK has wormhole like correlation given by non-zero value of $G_{LR}$ in the bulk. The quantum mechanical model with fixed choice of coupling \eqref{bm} gives exactly the same value of $G_{LR}$ when the holonomy of the gauge group is identity. 
\item The SYK model which fixed choice of coupling has wormhole saddle point in the self-averaging regime which has weak dependence on the coupling similarly the tensor model also has self-averaging wormhole saddle points described by trivial eigenvalues of gauge holonomy.
\item The SYK model with fixed choice of coupling has half-wormhole saddle points in the non self-averaging regime of the theory which are necessary for the factorization of the partition function. However, in the tensor model the dynamics of the gauge holonomy some how give rise to these half-wormhole saddle points which are in turn described by there non-trivial eigenvalues.
\item The SFF of SYK has linear growth of ramp w.r.t time and at late times as fluctuations of the height of ramp when we consider fixed choice of coupling. However the SFF of tensor model (with holonomy contribution) particularly in the high temperature regime has a plateau like structure which comes from the trivial saddle points of holonomy and then has an exponential rise dominated by the non-trivial saddle points around $u=1$. 
\end{itemize}
\section{\textcolor{Sepia}{\textbf{ \Large Conclusion}}}\label{CC}
In this paper, we have given an interpretation of half-wormholes in the bulk with gauge field. These half-wormholes were discussed in the SYK model with fixed coupling as a saddle point of $\Phi(\sigma)$ near $\sigma = 0$ which vanish when we take average over the ensemble. These half-wormhole play an important role in restoring the factorization. We have shown how dynamics of the gauge holonomy from the decoupled L and R system contributes to the wormhole like correlation between boundary partition function as in SYK. We discussed how the problem of factorization of the Wilson operator at short distances could be transformed as a problem of global charges and non-trivial cobordism classes associated with the disconnected wormhole. In order to break global symmetry these wormholes needs to have defect which cancel the global charges and produces new cobordism classes which were not previously connected. The new wormholes with added topological defect is what we interpret as half-wormhole which are necessary for factorization. It will be interesting to know if one could think of these defects as perturbation around the wormhole as discussed in  \cite{Mukhametzhanov:2021nea} to get the factorized answer of the square of the partition function in the SYK. We also commented on the behaviour of Spectral form factor w.r.t various parameters, particularly w.r.t temperature we found that at low temperatures the half-wormhole saddle contributes to the SFF whereas at high temperatures the wormhole saddle play an important part. Last but not the least,  we have explicitly studied the ensemble averaged energetics in terms of energy, free energy and entropy from $O(N)^{q-1}$ tensor model summing up all the possible contributions coming from wormhole saddle point and half-worm saddle points.  From this energetics study we have found that in the final results the wormhole saddle will not at all contribute from the present computations.  On the other hand,  all the contributions coming from the half-worm saddle points no-trivially contribute in the final results and explicitly shows the inverse temperature $\beta=1/T$ dependence in the large $N$ limiting results for the various measures of the energetics.  \\

The future prospects of the work are as follows:
\begin{itemize}
\item We know that dynamics of holonomy is described by the simple  effective action $S_{eff}(U)$ and by summing over infinite class of graph captures the leading deviations or perturbation away from free Hamiltonian $H_{0}$. It would interesting to know that the dynamics of holonomy which give rise to "half-wormhole" saddle points in the bulk can also be seen as a hint of perturbation around wormhole in the $O(N)^{q-1}$ tensor model.

\item We have described wormholes with topological defects as the half-wormholes in which factorizes the Wilson operator at short distances. One can now ask what kind of cobordism classes comes into play which breaks the global symmetry?

\item The out of time order correlator (OTOC) has been calculated in the bulk dual of regular SYK. One can now ask how the dynamics of gauge holonomy contributes in the calculations of (OTOC) in the bulk.

\end{itemize}

\textbf{Acknowledgement:}
~~~The research fellowship of SC is supported by the J.  C.  Bose National Fellowship of Sudhakar Panda.  SC also would line to thank School of Physical Sciences, National Institute for Science Education and Research (NISER),  Bhubaneswar for providing the work friendly environment.  SC also thank all the members of our newly formed virtual international non-profit consortium Quantum Structures of the Space-Time \& Matter (QASTM) for elaborative discussions.  K. Shirish would like to thank VNIT Nagpur respectively, for providing fellowships.  
Last but not least,  we would like to acknowledge our debt to the people belonging to the various part of the world for their generous and steady support for research in natural sciences.
\\ \\
\textbf{Corresponding author address:}\\
E-mail:~sayantan.choudhury@niser.ac.in,  \\
$~~~~~~~~~~~~$sayanphysicsisi@gmail.com

\bibliographystyle{unsrt}

\providecommand{\href}[2]{#2}\begingroup\raggedright\begin{thebibliography}{}

\end{thebibliography}\endgroup


\begin{thebibliography}{99}

\bibitem{Page:1993wv}
D.~N.~Page,
``Information in black hole radiation,''
Phys. Rev. Lett. \textbf{71}, 3743-3746 (1993)
[\href{https://arxiv.org/abs/hep-th/9306083}{\ttfamily 	arXiv:hep-th/9306083}]

\bibitem{Almheiri:2019qdq}
A.~Almheiri, T.~Hartman, J.~Maldacena, E.~Shaghoulian and A.~Tajdini,
``Replica Wormholes and the Entropy of Hawking Radiation,''
JHEP \textbf{05}, 013 (2020)
[\href{https://arxiv.org/abs/1911.12333}{\ttfamily arXiv:1911.12333[hep-th]}]

\bibitem{Almheiri:2019hni}
A.~Almheiri, R.~Mahajan, J.~Maldacena and Y.~Zhao,
``The Page curve of Hawking radiation from semiclassical geometry,''
JHEP \textbf{03}, 149 (2020)
[\href{https://arxiv.org/abs/1908.10996}{\ttfamily arXiv:1908.10996[hep-th]}]

\bibitem{Maldacena:2018lmt}
J.~Maldacena and X.~L.~Qi,
``Eternal traversable wormhole,''
[\href{https://arxiv.org/abs/1804.00491 }{\ttfamily arXiv:1804.00491 [hep-th]}]

\bibitem{Penington:2019kki}
G.~Penington, S.~H.~Shenker, D.~Stanford and Z.~Yang,
``Replica wormholes and the black hole interior,''
[\href{https://arxiv.org/abs/1911.11977 }{\ttfamily arXiv:1911.11977[hep-th]}]

\bibitem{Maldacena:2017axo}
J.~Maldacena, D.~Stanford and Z.~Yang,
``Diving into traversable wormholes,''
 Fortsch. Phys. \textbf{65}, no.5, 1700034 (2017)
[\href{https://arxiv.org/abs/1704.05333 }{\ttfamily arXiv:1704.05333[hep-th]}]

\bibitem{Maldacena:2004rf}
J.~M.~Maldacena and L.~Maoz,
``Wormholes in AdS,''
JHEP \textbf{02}, 053 (2004)
[\href{https://arxiv.org/abs/0401024 }{\ttfamily arXiv:0401024[hep-th]}]

\bibitem{Marolf:2021kjc}
D.~Marolf and J.~E.~Santos,
``AdS Euclidean wormholes,''
[\href{https://arxiv.org/abs/2101.08875 }{\ttfamily arXiv:2101.08875[hep-th]}]

\bibitem{Johnson:2021rsh}
C.~V.~Johnson,
``On the Quenched Free Energy of JT Gravity and Supergravity,''
[\href{https://arxiv.org/abs/2104.02733 }{\ttfamily arXiv:2104.02733[hep-th]}]

\bibitem{Stanford:2020wkf}
D.~Stanford,
``More quantum noise from wormholes,''
[\href{https://arxiv.org/abs/2008.08570 }{\ttfamily arXiv:2008.08570[hep-th]}]

\bibitem{Saad:2021rcu}
P.~Saad, S.~H.~Shenker, D.~Stanford and S.~Yao,
``Wormholes without averaging,''
[\href{https://arxiv.org/abs/2103.16754}{\ttfamily arXiv:2103.16754 [hep-th]}]

\bibitem{kk}
Talk on "Wormholes and microstructure" by Stephen Shenker at Microstate Conference 2021,  organized by String Theory in Greater Paris,~
\href{https://www.youtube.com/watch?v=yQ0Q58FvsHM
Microstate conference 2021}{\ttfamily https://www.youtube.com/watch?v=yQ0Q58FvsHM
Microstate conference 2021}.

\bibitem{ramsey}
Talk on "Wormholes without averaging" by Douglas Stanford at Rencontres, organized by String Theory in Greater Paris,~
\href{https://www.youtube.com/watch?v=TMjZvbBYfz0
String theory Talk 15.4.2021}{\ttfamily https://www.youtube.com/watch?v=TMjZvbBYfz0
String theory Talk 15.4.2021}.

\bibitem{philsaadtalk}
Talk on "Some comments on wormholes and factorization" by Phil Saad at Microstate Conference 2021,  organized by String Theory in Greater Paris,~
\href{https://www.youtube.com/watch?v=zB99h7iWKOg}{\ttfamily https://www.youtube.com/watch?v=zB99h7iWKOg}.

\bibitem{LR}
Talk on "Black holes, wormholes, long times and ensemble" by Stephen Shenker at Strings 2021,  Brazil,~
\href{https://youtu.be/vDj3uI2vsck
ICTP-SAIFR Strings 2021}{\ttfamily https://youtu.be/vDj3uI2vsck
ICTP-SAIFR Strings 2021}.

\bibitem{Cao:2021upq}
S.~Cao, Y.~C.~Rui and X.~H.~Ge,
``Thermodynamic phase structure of complex Sachdev-Ye-Kitaev model and charged black hole in deformed JT gravity,''
[\href{https://arxiv.org/abs/2103.16270}{\ttfamily arXiv:2103.16270 [hep-th]}]

\bibitem{Sachdev:1992fk}
S.~Sachdev and J.~Ye,
``Gapless spin fluid ground state in a random, quantum Heisenberg magnet,''
Phys. Rev. Lett. \textbf{70}, 3339 (1993)
[\href{https://arxiv.org/abs/9212030}{\ttfamily arXiv:9212030[hep-th]}]

\bibitem{Hosur:2015ylk}
P.~Hosur, X.~L.~Qi, D.~A.~Roberts and B.~Yoshida,
``Chaos in quantum channels,''
JHEP \textbf{02}, 004 (2016)
[\href{https://arxiv.org/abs/1511.04021}{\ttfamily arXiv:1511.04021[hep-th]}]

\bibitem{McNamara:2019rup}
J.~McNamara and C.~Vafa,
``Cobordism Classes and the Swampland,''
[\href{https://arxiv.org/abs/1909.10355 }{\ttfamily arXiv:1909.10355 [hep-th]}]

\bibitem{Jevicki:2016bwu}
A.~Jevicki, K.~Suzuki and J.~Yoon,
``Bi-Local Holography in the SYK Model,''
JHEP \textbf{07}, 007 (2016)
[\href{https://arxiv.org/abs/1603.06246}{\ttfamily arXiv:1603.06246[hep-th]}]

\bibitem{Polchinski:2016xgd}
J.~Polchinski and V.~Rosenhaus,
``The Spectrum in the Sachdev-Ye-Kitaev Model,''
JHEP \textbf{04}, 001 (2016)
[\href{https://arxiv.org/abs/1601.06768}{\ttfamily arXiv:1601.06768[hep-th]}]

\bibitem{Gross:2016kjj}
D.~J.~Gross and V.~Rosenhaus,
``A Generalization of Sachdev-Ye-Kitaev,''
JHEP \textbf{02}, 093 (2017)
[\href{https://arxiv.org/abs/1610.01569}{\ttfamily arXiv:1610.01569[hep-th]}]

\bibitem{Berkooz:2016cvq}
M.~Berkooz, P.~Narayan, M.~Rozali and J.~Sim\'on,
``Higher Dimensional Generalizations of the SYK Model,''
JHEP \textbf{01}, 138 (2017)
[\href{https://arxiv.org/abs/1610.02422}{\ttfamily arXiv:1610.02422[hep-th]}]


\bibitem{Maldacena:2016hyu}
J.~Maldacena and D.~Stanford,
``Remarks on the Sachdev-Ye-Kitaev model,''
Phys. Rev. D \textbf{94}, no.10, 106002 (2016)
[\href{https://arxiv.org/abs/1604.07818}{\ttfamily arXiv:1604.07818[hep-th]}]

\bibitem{Gu:2016oyy}
Y.~Gu, X.~L.~Qi and D.~Stanford,
``Local criticality, diffusion and chaos in generalized Sachdev-Ye-Kitaev models,''
JHEP \textbf{05}, 125 (2017)
[\href{https://arxiv.org/abs/1609.07832 }{\ttfamily arXiv:1609.07832 [hep-th]}]

\bibitem{Saad:2018bqo}
P.~Saad, S.~H.~Shenker and D.~Stanford,
``A semiclassical ramp in SYK and in gravity,''
[\href{https://arxiv.org/abs/1806.06840}{\ttfamily arXiv:1806.06840 [hep-th]}]

\bibitem{You:2016ldz}
Y.~Z.~You, A.~W.~W.~Ludwig and C.~Xu,
``Sachdev-Ye-Kitaev Model and Thermalization on the Boundary of Many-Body Localized Fermionic Symmetry Protected Topological States,''
Phys. Rev. B \textbf{95}, no.11, 115150 (2017)
[\href{https://arxiv.org/abs/1602.06964}{\ttfamily arXiv:1602.06964 [hep-th]}]


\bibitem{Sachdev:2010um}
S.~Sachdev,
``Holographic metals and the fractionalized Fermi liquid,''
Phys. Rev. Lett. \textbf{105}, 151602 (2010)
[\href{https://arxiv.org/abs/1006.3794}{\ttfamily arXiv:1006.3794 [hep-th]}]

\bibitem{Freivogel:2021ivu}
B.~Freivogel, D.~Nikolakopoulou and A.~F.~Rotundo,
``Wormholes from Averaging over States,''
[\href{https://arxiv.org/abs/2105.12771}{\ttfamily arXiv:2105.12771[hep-th]}]

\bibitem{Berkooz:2021ewq}
M.~Berkooz, N.~Brukner, V.~Narovlansky and A.~Raz,
``Multitrace Correlators in the SYK Model and Non-geometric Wormholes,''
[\href{https://arxiv.org/abs/2104.03336}{\ttfamily arXiv:2104.03336[hep-th]}]

\bibitem{Choudhury:2017tax}
S.~Choudhury, A.~Dey, I.~Halder, L.~Janagal, S.~Minwalla and R.~Poojary,
``Notes on melonic $O(N)^{q-1}$ tensor models,''
JHEP \textbf{06}, 094 (2018)
[\href{https://arxiv.org/abs/1707.09352}{\ttfamily arXiv:1707.09352[hep-th]}]

\bibitem{Johnson:2020heh}
C.~V.~Johnson,
``Jackiw-Teitelboim supergravity, minimal strings, and matrix models,''
Phys. Rev. D \textbf{103}, no.4, 046012 (2021)
[\href{https://arxiv.org/abs/2005.01893}{\ttfamily arXiv:2005.01893[hep-th]}]

\bibitem{Gurau:2010ba}
R.~Gurau,
``The 1/N expansion of colored tensor models,''
Annales Henri Poincare \textbf{12}, 829-847 (2011)
[\href{https://arxiv.org/abs/1011.2726}{\ttfamily arXiv:1011.2726[hep-th]}]

\bibitem{Gurau:2011aq}
R.~Gurau and V.~Rivasseau,
``The 1/N expansion of colored tensor models in arbitrary dimension,''
EPL \textbf{95}, no.5, 50004 (2011)
[\href{https://arxiv.org/abs/1101.4182}{\ttfamily arXiv:1101.4182[hep-th]}]

\bibitem{Gurau:2011xq}
R.~Gurau,
``The complete 1/N expansion of colored tensor models in arbitrary dimension,''
Annales Henri Poincare \textbf{13}, 399-423 (2012)
[\href{https://arxiv.org/abs/1102.5759}{\ttfamily arXiv:1102.5759[hep-th]}]

\bibitem{Bonzom:2011zz}
V.~Bonzom, R.~Gurau, A.~Riello and V.~Rivasseau,
``Critical behavior of colored tensor models in the large N limit,''
Nucl. Phys. B \textbf{853}, 174-195 (2011)
[\href{https://arxiv.org/abs/1105.3122}{\ttfamily arXiv:1105.3122[hep-th]}]

\bibitem{Bonzom:2012hw}
V.~Bonzom, R.~Gurau and V.~Rivasseau,
``Random tensor models in the large N limit: Uncoloring the colored tensor models,''
Phys. Rev. D \textbf{85}, 084037 (2012)
[\href{https://arxiv.org/abs/1202.3637}{\ttfamily arXiv:1202.3637[hep-th]}]

\bibitem{Sasakura:1990fs}
N.~Sasakura,
``Tensor model for gravity and orientability of manifold,''
Mod. Phys. Lett. A \textbf{6}, 2613-2624 (1991)


\bibitem{Witten:2016iux}
E.~Witten,
``An SYK-Like Model Without Disorder,''
J. Phys. A \textbf{52}, no.47, 474002 (2019)
[\href{https://arxiv.org/abs/1610.09758}{\ttfamily arXiv:1610.09758[hep-th]}]

\bibitem{Gurau:2016cjo}
R.~Gurau,
SIGMA \textbf{12}, 094 (2016)
[\href{https://arxiv.org/abs/1609.06439}{\ttfamily arXiv:1609.06439[hep-th]}]

\bibitem{Harlow:2015lma}
D.~Harlow,
``Wormholes, Emergent Gauge Fields, and the Weak Gravity Conjecture,''
JHEP \textbf{01}, 122 (2016)
[\href{https://arxiv.org/abs/1510.07911}{\ttfamily arXiv:1510.07911[hep-th]}]

\bibitem{Hamilton:2006az}
A.~Hamilton, D.~N.~Kabat, G.~Lifschytz and D.~A.~Lowe,
``Holographic representation of local bulk operators,''
Phys. Rev. D \textbf{74}, 066009 (2006)
[\href{https://arxiv.org/abs/0606141}{\ttfamily arXiv:0606141[hep-th]}]

\bibitem{Kabat:2011rz}
D.~Kabat, G.~Lifschytz and D.~A.~Lowe,
``Constructing local bulk observables in interacting AdS/CFT,''
Phys. Rev. D \textbf{83}, 106009 (2011)
[\href{https://arxiv.org/abs/1102.2910}{\ttfamily arXiv:1102.2910[hep-th]}]

\bibitem{Kabat:2013wga}
D.~Kabat and G.~Lifschytz,
``Decoding the hologram: Scalar fields interacting with gravity,''
Phys. Rev. D \textbf{89}, no.6, 066010 (2014)
[\href{https://arxiv.org/abs/1311.3020}{\ttfamily arXiv:1311.3020[hep-th]}]

\bibitem{ss}
Some comments on wormholes and factorization
https://www.youtube.com/watch?v=zB99h7iWKOg
Black-Hole Microstructure conference june 9, 2020



\bibitem{Gaiotto:2014kfa}
D.~Gaiotto, A.~Kapustin, N.~Seiberg and B.~Willett,
``Generalized Global Symmetries,''
JHEP \textbf{02}, 172 (2015)
[\href{https://arxiv.org/abs/1412.5148}{\ttfamily arXiv:1412.5148[hep-th]}]

\bibitem{Harlow:2018tng}
D.~Harlow and H.~Ooguri,
``Symmetries in quantum field theory and quantum gravity,''
Commun. Math. Phys. \textbf{383}, no.3, 1669-1804 (2021)
[\href{https://arxiv.org/abs/1810.05338}{\ttfamily arXiv:1810.05338[hep-th]}]

\bibitem{Banks:2010zn}
T.~Banks and N.~Seiberg,
``Symmetries and Strings in Field Theory and Gravity,''
Phys. Rev. D \textbf{83}, 084019 (2011)
[\href{https://arxiv.org/abs/1011.5120}{\ttfamily arXiv:1011.5120[hep-th]}]

\bibitem{Distler:2009ri}
J.~Distler, D.~S.~Freed and G.~W.~Moore,
``Orientifold Precis,''
[\href{https://arxiv.org/abs/0906.0795}{\ttfamily arXiv:0906.0795[hep-th]}]

\bibitem{Belin:2020jxr}
A.~Belin, J.~De Boer, P.~Nayak and J.~Sonner,
``Charged Eigenstate Thermalization, Euclidean Wormholes and Global Symmetries in Quantum Gravity,''
[\href{https://arxiv.org/abs/2012.07875}{\ttfamily arXiv:2012.07875[hep-th]}]

\bibitem{DAlessio:2016rwt}
L.~D'Alessio, Y.~Kafri, A.~Polkovnikov and M.~Rigol,
``From quantum chaos and eigenstate thermalization to statistical mechanics and thermodynamics,''
Adv. Phys. \textbf{65}, no.3, 239-362 (2016)
[\href{https://arxiv.org/abs/1509.06411}{\ttfamily arXiv:1509.06411[hep-th]}]

\bibitem{Balasubramanian:2011ur}
V.~Balasubramanian, A.~Bernamonti, J.~de Boer, N.~Copland, B.~Craps, E.~Keski-Vakkuri, B.~Muller, A.~Schafer, M.~Shigemori and W.~Staessens,
``Holographic Thermalization,''
Phys. Rev. D \textbf{84}, 026010 (2011)
[\href{https://arxiv.org/abs/1103.2683}{\ttfamily arXiv:1103.2683[hep-th]}]

\bibitem{Sonner:2017hxc}
J.~Sonner and M.~Vielma,
``Eigenstate thermalization in the Sachdev-Ye-Kitaev model,''
JHEP \textbf{11}, 149 (2017)
[\href{https://arxiv.org/abs/1707.08013}{\ttfamily arXiv:1707.08013[hep-th]}]

\bibitem{Anous:2019yku}
T.~Anous and J.~Sonner,
``Phases of scrambling in eigenstates,''
SciPost Phys. \textbf{7}, 003 (2019)
[\href{https://arxiv.org/abs/1903.03143}{\ttfamily arXiv:1903.03143[hep-th]}]

\bibitem{Garrison:2015lva}
J.~R.~Garrison and T.~Grover,
``Does a single eigenstate encode the full Hamiltonian?,''
Phys. Rev. X \textbf{8}, no.2, 021026 (2018)
[\href{https://arxiv.org/abs/1503.00729}{\ttfamily arXiv:1503.00729[hep-th]}]

\bibitem{Susskind:2021omt}
L.~Susskind,
``De Sitter Holography: Fluctuations, Anomalous Symmetry, and Wormholes,''
[\href{https://arxiv.org/abs/2106.03964}{\ttfamily arXiv:2106.03964[hep-th]}]

\bibitem{Kapustin:2006pk}
A.~Kapustin and E.~Witten,
``Electric-Magnetic Duality And The Geometric Langlands Program,''
Commun. Num. Theor. Phys. \textbf{1}, 1-236 (2007)
[\href{https://arxiv.org/abs/0604151}{\ttfamily arXiv:0604151[hep-th]}]

\bibitem{Gaikwad:2017odv}
A.~Gaikwad and R.~Sinha,
``Spectral Form Factor in Non-Gaussian Random Matrix Theories,''
Phys. Rev. D \textbf{100}, no.2, 026017 (2019)
[\href{https://arxiv.org/abs/1706.07439}{\ttfamily arXiv:1706.07439[hep-th]}]

\bibitem{Choudhury:2018lcb}
S.~Choudhury and A.~Mukherjee,
``A bound on quantum chaos from Random Matrix Theory with Gaussian Unitary Ensemble,''
JHEP \textbf{05}, 149 (2019)
[\href{https://arxiv.org/abs/1811.01079}{\ttfamily arXiv:1811.01079[hep-th]}]

\bibitem{Choudhury:2018rjl}
S.~Choudhury, A.~Mukherjee, P.~Chauhan and S.~Bhattacherjee,
``Quantum Out-of-Equilibrium Cosmology,''
Eur. Phys. J. C \textbf{79}, no.4, 320 (2019)
[\href{https://arxiv.org/abs/1809.02732}{\ttfamily arXiv:1809.02732[hep-th]}]

\bibitem{Grassi:2019txd}
A.~Grassi, Z.~Komargodski and L.~Tizzano,
``Extremal correlators and random matrix theory,''
JHEP \textbf{04}, 214 (2021)
[\href{https://arxiv.org/abs/1908.10306}{\ttfamily arXiv:1908.10306[hep-th]}]

\bibitem{Guhr:1997ve}
T.~Guhr, A.~Muller-Groeling and H.~A.~Weidenmuller,
``Random matrix theories in quantum physics: Common concepts,''
Phys. Rept. \textbf{299}, 189-425 (1998)
[\href{https://arxiv.org/abs/9707301}{\ttfamily arXiv:9707301[hep-th]}]
\bibitem{Cotler:2017jue}
J.~Cotler, N.~Hunter-Jones, J.~Liu and B.~Yoshida,
``Chaos, Complexity, and Random Matrices,''
JHEP \textbf{11}, 048 (2017)
[\href{https://arxiv.org/abs/1706.05400}{\ttfamily arXiv:1706.05400[hep-th]}]

\bibitem{Kodama:2008zra}
Y.~Kodama and V.~U.~Pierce,
``Combinatorics of dispersionless integrable systems and universality in random matrix theory,''
Commun. Math. Phys. \textbf{292}, 529 (2009)
[\href{https://arxiv.org/abs/0811.0351}{\ttfamily arXiv:0811.0351[hep-th]}]

\bibitem{Cotler:2016fpe}
J.~S.~Cotler, G.~Gur-Ari, M.~Hanada, J.~Polchinski, P.~Saad, S.~H.~Shenker, D.~Stanford, A.~Streicher and M.~Tezuka,
``Black Holes and Random Matrices,''
JHEP \textbf{05}, 118 (2017)
[erratum: JHEP \textbf{09}, 002 (2018)]
[\href{https://arxiv.org/abs/1611.04650}{\ttfamily arXiv:1611.04650[hep-th]}]

\bibitem{Mukhametzhanov:2021nea}
B.~Mukhametzhanov,
``Half-wormhole in SYK with one time point,''
[\href{https://arxiv.org/abs/2105.08207}{\ttfamily arXiv:2105.08207[hep-th]}]
\end{thebibliography}

\end{document}